\documentclass[12pt,preprint]{aastex}


\begin{document}


\title{Probing The Cosmological Constant Through The Alcock-Paczynski Test Based on The Lyman-Alpha Forest}

\author{Wen-Ching Lin\altaffilmark{1,2,3} and Michael L. Norman \altaffilmark{1,2}}





\altaffiltext{1}{Physics Department, University of California at San Diego, San Diego, CA 92093}
\altaffiltext{2}{Center for Astrophysics and Space Sciences, University of California at San Diego, San Diego, CA 92093}
\altaffiltext{3}{Astronomy Department, University of Illinois, Urbana, IL 61801}


\begin{abstract}
  In recent years, the possibility of measuring the cosmological constant $\Omega_\Lambda$ through the application of the Alcock-Paczynski test to the Lyman Alpha (Ly$\alpha$) forest has been suggested (McDonald et al. 1999; Hui et al. 1999). Despite the theoretical uncertainties due to a few other cosmological parameters, some of the greatest difficulties we encounter concern the huge uncertainties due to cosmic variance and noise. In this paper, we propose a maximum likelihood estimation (MLE) method to deal with cosmic variance and noise using synthetic spectra of quasistellar objects (QSOs) from our cosmological hydrodynamic simulations. We demonstrate that the MLE method can overcome the cosmic variance problem. Applying the MLE method, we find that we have more than 90$\%$ probability to determine $\Omega_\Lambda$ within 20$\%$ error and approximately of 66$\%$ probability to determine  $\Omega_\Lambda$ within 10$\%$ error  by using 30 pairs QSO spectra when other cosmological parameters are assumed. Another important source of error is from noise in the flux spectra, and we have modeled the corresponding effect by studying artificial spectra with different kinds of noise added. We discover that the noise distribution does not have significant effect on the final cross-correlation functions as long as the signal-to-noise ratio (S/N) is fixed. Finally, a preliminary test and discussion about the sensitivities to other cosmological parameters are included in this paper as well.

\end{abstract}

\keywords{Ly$\alpha$ forest, cosmological constant, AP test}

\section{Introduction}
\label{sec:APT}

An important topic in cosmology is the determination of the energy densities of the various components of the Universe. Therefore constraining the values of their respective fractional energy densities: baryon density ($\Omega_{B}$), matter density ($\Omega_{M}$) and vacuum density ($\Omega_{\Lambda}$) becomes a vital part of understanding the Universe. First year WMAP results gives $\Omega_0=$1.02$\pm$0.02 (Bennett et al. 2003).
When combined with galaxy clustering data, the Lyman $\alpha$ forest, and other CMB measurements, WMAP finds $\Omega_B h^2$=0.0224$\pm$0.0009 and 
$\Omega_M$=0.268$\pm$0.0159 (Spergel et al. 2003).

The results $\Omega_{0} \approx 1$ and $\Omega_{M}\approx 0.27$ suggest the existence of a dark, exotic form of energy, which is smoothly distributed and contributes roughly 70$\%$ of the critical density (Turner et al, 1983; Peebles, 1984). It is summarized $\Omega_{\Lambda}$ = $\frac{4}{3}$ $\Omega_{M}$ + $\frac{1}{3}$ $\pm$ $\frac{1}{6}$ based on observations of the Type Ia supernovae (Perlmutter et al. 1999; Goobar 2000; Riess et al. 1998; Schmidt et al. 1998), which is consistent with studies based on different physics (Holder et al. 2000; Guerra et al. 2000). Since all these results are based on data at z $<$ 2 (mostly $<$ 1), other independent measurements of $\Omega_{\Lambda}$ from data at an earlier epoch becomes important. This motivates us to implement the Alcock-Paczynski Test (AP test) on the Lyman Alpha (Ly$\alpha$) forest.

 In 1979, Alcock  and Paczynski proposed a method which can be used to measure the geometry of the Universe (Alcock $\&$ Paczynski, 1979). The basic idea of this method is that for a spherical object in the sky, its physical size along the line of sight and perpendicular to the line of sight should be equal. This method can be extended to non-spherical cosmological structures, in which case the characteristic length of the correlation function of the structure will be used instead of the physical size of the object. The two-point correlation functions of galaxies and clusters have been suggested as candidates for the AP test (Ryden 1995; Ballinger et al. 1996; Matsubara $\&$ Suto 1996; Popowski et al. 1998). Recently the cross-correlation function of Ly$\alpha$ forest clouds has been proposed as a good candidate for the AP test as well (McDonald et al. 1999; Hui et al. 1999; McDonald 2003).

   The geometrical basis of the AP test suggests the measurement of the value $\frac{\Delta z}{z \Delta \theta}$, where $\frac{\Delta z}{z}$ and $\Delta \theta$ relate to the scales of the object due to the Hubble flow expansion parallel and perpendicular to the line of sight, respectively. Observationally, the characteristic lengths parallel and perpendicular to the line of sight can be written in terms of the velocity separations $v_{\Vert}$ and $v_{\perp}$. When $\Delta$z is small, the velocity parallel to the line of sight is given by :

\begin{equation}
 \Delta v_{\Vert} = \frac{\Delta z}{1+z} c = \Delta v_{h} + \Delta v_{p}
\label{eqn:v_para.eqn}
\end{equation}

 where $\Delta v_{h}$ is the velocity separation due to the Hubble flow expansion and $\Delta v_{p}$ denotes the effect of the peculiar velocity. The transverse velocity separation is :

\begin{equation}
 \Delta v_{\perp} = H(z)  \Delta l = H(z)  D_{A}(z)  \Delta  \theta
\label{eqn:v_perp.eqn}
\end{equation}
 where $D_{A}$(z) is the redshift-dependent angular diameter distance, H(z) is the expansion rate and $\Delta l$ is the physical size of the object we are interested in. We can write $\Delta v_{\perp}$ in a form expressing its relationship with cosmological parameters explicitly : $\Delta v_{\perp}$ = c f(z) $\Delta  \theta$, where

\begin{equation}
 f(z) = \frac{1}{c} D_{A}(z) H(z) = \frac{E(z)}{1+z} \int_{0}^{z} \frac{dz'}{E(z')}
\label{eqn:fz.eqn}
\end{equation}

 with E(z) = $[\Omega_{m} (1+z)^{3} + \Omega_{\Lambda}(1+z)^{3(1+\omega)}] ^{\frac{1}{2}}$ where we specialize to the case with an equation of state $\omega \equiv \frac{dp}{d \rho} $ = -1.

 Because the matter distribution expands with the Hubble flow at the same rate in all directions in a homogeneous and isotropic universe, the formation of structures should be equal in all directions statistically. Therefore, the auto-correlation function along the line of sight $\xi$($\Delta$ $v_\parallel$) and the auto-correlation function perpendicular to the line of sight $\xi$($\Delta$ $v_\perp$) should be the same (see $\S$\ref{subsec:CrossCorre} for detailed definitions of correlation functions). Unfortunately, one can not observe the auto-correlation function perpendicular to the line of sight. However, it is suggested that one can overcome this problem by analytically deriving a cross-correlation function $\xi_\times$($\Delta$ $v_\parallel$) = $\xi_\times$( $\sqrt{\Delta v^2 - \Delta v_\perp^2}$) of two parallel lines of sight based on the information from the auto-correlation function. Then by comparing with the observational cross-correlation function, we can determine the cosmological model (McDonald et al. 1999; Hui et al. 1999). Based on this idea, instead of analytically deriving cross-correlation functions, we use fully hydrodynamical simulations to provide more accurate results. 
The benefit of this approach is that the nonlinear effects (e.g., peculiar velocities, shock heating) are automatically included. 
We also able to generate many numerical paired QSO spectra for the purpose of statistical analysis and the comparison with real observational data.

\begin{figure}
\plotone{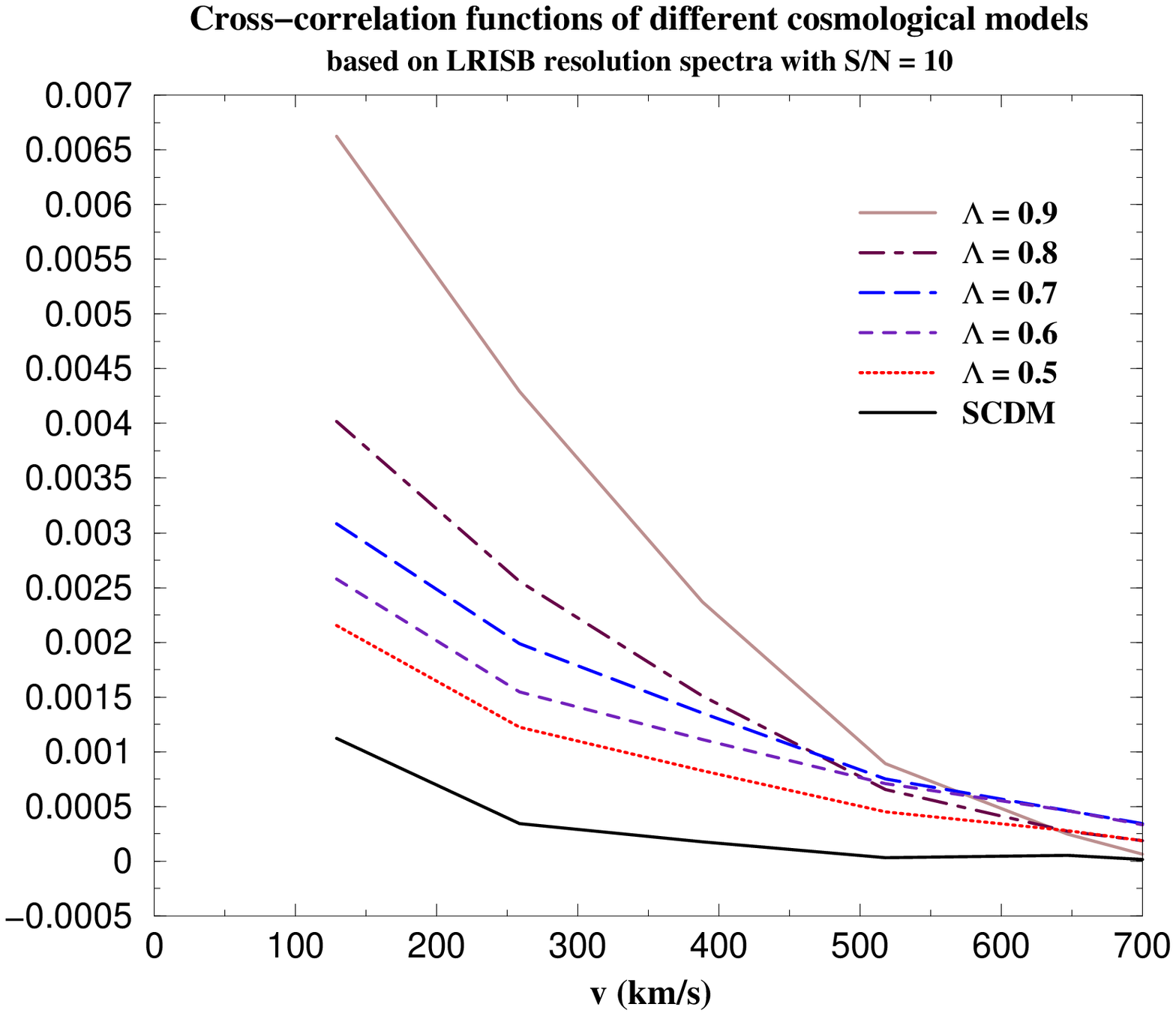}
\caption{Flux cross-correlation functions for several cosmological models with different $\Omega_\Lambda$. Each curve is the average of 300 cross-correlation functions while each cross-correlation function was derived from a paired QSO spectra based on our simulations. The corresponding FWHM is 300 km/s, the pixel size is 130 km/s and the signal-to-noise ratio is 10. The angular separation is 120". The lower solid curve is the SCDM model and the higher solid curve is the $\Omega_\Lambda$ = 0.9 model. The dotted curve is the $\Omega_\Lambda$ = 0.5 model, the dashed curve is the $\Omega_\Lambda$ = 0.6 model, the long-dashed curve is the $\Omega_\Lambda$ = 0.7 model and the dotted-dashed curve is the $\Omega_\Lambda$ = 0.8 model.}
\label{fig:CrossCorre.fig}
\end{figure}

 
Figure (\ref{fig:CrossCorre.fig}) shows the flux cross-correlation functions in different cosmological models as taken from our simulations. 
As is evident, the flux cross-correlation function is a powerful descriminant between models. 
In the specific structures we are interested in, the Ly$\alpha$ forest at redshift 2, the velocity separations of 100 - 600 km/s correspond to comoving 
scales of 1 - 6 Mpc, which is the characteristic size of voids lying between the absorbing clouds.  
Within the current paradign (e.g., Zhang et al. 1998), Ly$\alpha$ forest
absorbers are mildly overdense regions of gas that form a network of 
sheets and filaments that are nearly fixed in comoving coordinates.
This makes them perfect candidates for the AP test. Complications
due to nonlinear processes (shock heating, peculiar velocities)
are taken into account in our analysis since they are included in
the underlying simulations. 

This paper is the first of several describing our methodology for
applying the AP test to the Lyman alpha forest. Here we focus on the
uncertainties of estimating $\Omega_\Lambda$ due to cosmic variance
and noise in the quasar spectra assuming all other parameters are
known. We describe a method, based on maximum liklihood estimation
(MLE), that overcomes these problems given a sufficient number of
quasar pairs. We show that uncertainties due to the cosmological
parameters $\Omega_b$ and $\sigma_8$ are small, while the uncertainties
due to the poorly constrained intensity of the UV background are of the 
same order as uncertainties due to $\Omega_\Lambda$. This latter
issue is currently under investigation by us and will be reported on
in a forthcoming paper. 

This paper is organized as follows : The cosmological simulations and the analysis codes are discussed in $\S$\ref{sec:CosSim} and $\S$\ref{sec:SimSpec}. Then we display our methodology in $\S$\ref{sec:Method}. Finally, we provide our results and concluding remarks in $\S$\ref{sec:Res} and $\S$\ref{sec:DisAndCon}. Additional information about maximum likelihood estimation and error propagation are given in Appendix (\ref{sec:MLE}) and Appendix (\ref{subsec:ErrPro}).

\section{Cosmological Simulations}
\label{sec:CosSim}
\subsection{Cosmological Code}
\label{sec:CosCode}
  We have performed several simulations of the z = 2 Ly$\alpha$ forest in different cosmological models. All simulations were performed using our cosmological hydrodynamics code Enzo. Enzo incorporates a Lagrangean particle-mesh (PM) algorithm to follow the collisionless dark matter and a higher-order accurate piecewise parabolic method (PPM) to solve the equations of gas dynamics. In addition to the usual ingredients of baryonic and dark matter, Enzo also solves a coupled system of non-equilibrium ionization equations with radiative cooling for a gas with primordial abundances. Our chemical reaction network includes six species: HI, HII, HeI, HeII, HeIII and $e^{-}$ (Abel et al. 1997; Anninos et al. 1997). The simulation starts with the initial perturbations originating from inflation-inspired adiabatic fluctuations. The BBKS (Bardeen et al. 1986) transfer function is employed with the standard Harrison-Zel'dovich power spectrum. Another important component in the simulations is an ultraviolet (UV) radiation background which ionizes the neutral intergalactic medium. Haardt $\&$ Madau (1996) have provided a UV radiation field with a radiation transfer model in a clumpy universe based upon the observed quasar luminosity function. Enzo starts to import their homogeneous UV background spectra at redshift 7 and increases the intensity of the spectra at redshift 6 to generate photoionization and photoheating rates in our simulations.

\subsection{Ly$\alpha$ Forest Simulations}
\label{subsec:LyaSim}
  In this work we performed twelve Ly$\alpha$ forest simulations using Enzo. All simulations were done on the Origin 2000 supercomputer at the National Center for Supercomputing Applications (NCSA). We have six major simulations for our main analysis work and their comoving box size are 37.3 Mpc (25.73 Mpc/h, h = 0.69) with $256^{3}$ dark matter particles and a $256^{3}$ grid for the evolution of gas dynamics. The other cosmological parameters used here are $\sigma_8$ = 0.73, $\Omega_B$ = 0.04, $\Omega_{0}$ = 1.0, h = 0.69 and the power spectrum index n = 1.0. The only parameter varied over the six simulations is $\Omega_\Lambda$, which is given values of 0.0, 0.5, 0.6, 0.7, 0.8 and 0.9. In addition to these, we have another six simulations with $\Omega_\Lambda$ = 0.7 where we varied other cosmological parameters: $\sigma_{8}$, photoionization parameter $J$ and $\Omega_{B}$(or h) for the purpose of testing parameter sensitivities. (see Table (\ref{table:CosPara.tbl}) and $\S$ \ref{subsec:SenCos} for a summary). 

\section{Artificial QSO spectra}
\label{sec:SimSpec}
\subsection{Spectrum Generator}
\label{subsec:SpeGen}
 In order to compare our simulation results to observations, we need to produce realistic artificial flux spectra from our simulations. The spectrum generator we used here starts at the point with the lowest neutral hydrogen density inside the box, shooting photons along random lines of sight through the box. Theoretically, one can start at any point in the box and it is just a matter of choice to choose the point with the lowest $N_{HI}$ density. Here we denote this spectrum generating method as SGA. Another spectrum generating method often used is to shoot lines of sight parallel to the edge of the box, and we denote this method as SGB. One important advantage of SGA is that it is actually closer to the real observational case. This is because when we observe the Universe, the observation always starts at a single point, Earth, and collects spectra from different lines of sight. Therefore, if this kind of observation does introduce statistical errors, we want to reproduce the same bias in the numerical simulations. The method SGA also avoids the dependence of nearby lines of sight in the method SGB. The method SGA has been used and carefully studied by Zhang et al. (1998),  Bryan et al. (1999), and Machacek et al. (2001).

 The method SGA calculates the transmitted flux of a QSO at redshift z as $e^{- \tau_{\nu}}$, with the optical depth $\tau_{\nu}$ given by

\begin{equation}
   \tau_{\nu} (t) \equiv \int_{t}^{t_{0}}   n_{HI}(t) \sigma_{\nu} c dt
\label{eqn:tau.eqn}
\end{equation}
  where c is the speed of light, $n_{HI}$ is the number density of the HI absorbers, $\sigma_{\nu}$ is the absorption cross-section, t is the corresponding cosmic time at redshift z and $t_{0}$ is the cosmic time today. Integration is performed along the line of sight from the QSO to the observer. This can be written in a form more suitable for computation (Zhang et.al. 1997) as

${ \tau_\nu(z) = \frac{c^2 \sigma_o}{\sqrt{\pi} \nu_o} \int_{z}^{z_o} 
   \frac{n_{HI}(\acute{z})}{b} \frac{a^2}{\dot{a}} 
   \,exp \left\{ - \left[ (1+\acute{z})\frac{\nu}{\nu_o} - 1 + \frac{v}{c} 
   \right]^{2} \frac{c^2}{b^2} \right\} d\!\acute{z}        }$

where $z$ is the redshift, $\sigma_o$ is the resonant Ly-${\alpha}$ cross
section,
 $\nu_o$ is the Ly$\alpha$ rest frequency, $v$ is the peculiar velocity along the line-of-sight and $\nu$ is the redshifted frequency. b is the effect of Doppler broadening on the absorption cross section and is equal to $\sqrt{2kT/m_p}$, where k is the Boltzmann's constant, T is the gas temperature and $m_p$ is the mass of a proton. This equation, parametrized to order ${v/c}$, also needs the scale factor $a$ to be specified, which is given by the Friedman equation,

${  \dot{a} = H_o \sqrt{1 + \Omega_m (\frac{1}{a}-1) + \Omega_{\Lambda} (a^2-1)} }$

\begin{figure}
\plotone{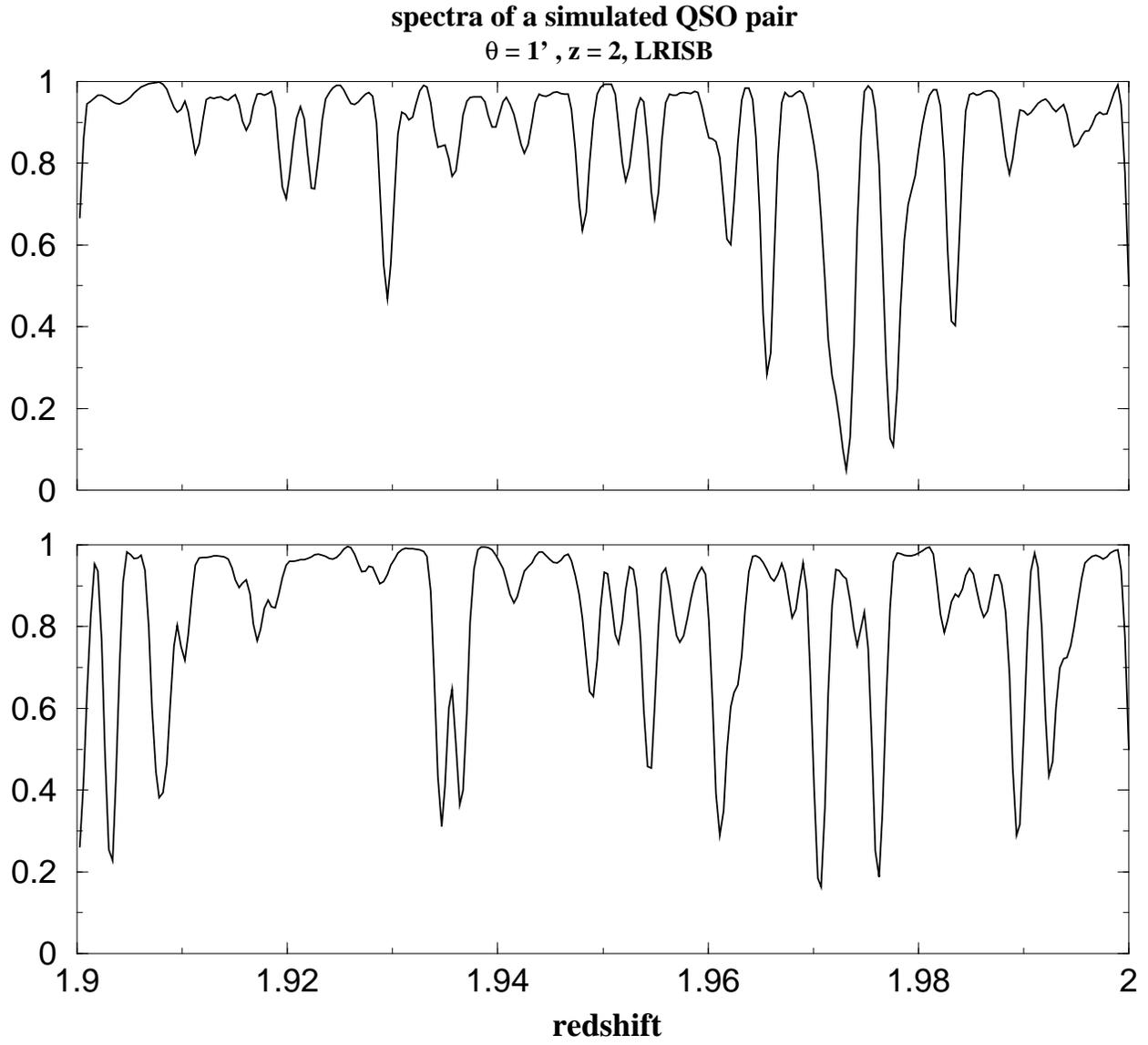}
\caption{A pair of simulated QSO spectra at z=2. The spectra were degraded to LRIS resolution with a FWHM of 300 km/s and a pixel size of 130 km/s. The angular separation of the QSO pair is 1'.}
\label{fig:Pair_LRISB.fig}
\end{figure}

  In order to generate paired QSO spectra, we first calculated the comoving separation d of two QSOs at the desired redshift based on the known angular separation under a given cosmological model. Then from the point with the lower neutral hydrogen density, point $A$, and a given random direction $\vec{r}$, the plane $S$ : ($\vec{x}$ - $\vec{A}$) $\cdot$ $\vec{r}$ = 0 is uniquely determined. Then on the circle with center $A$ and radius d/2 on $S$, we chose two points $C$ and $D$ where C-A-D lay on a line. Then we generated two parallel QSO spectra from $C$ and $D$ both along the direction $\vec{r}$. Figure (\ref{fig:Pair_LRISB.fig}) shows a pair of simulated QSO spectra with a resolution identical to the Low Resolution Imaging Spectrometer (LRIS) on the Keck telescope.

\subsection{Spectra Degrading and Signal to Noise Properties}
\label{subsec:SpeDeSN}
\begin{figure}
\plotone{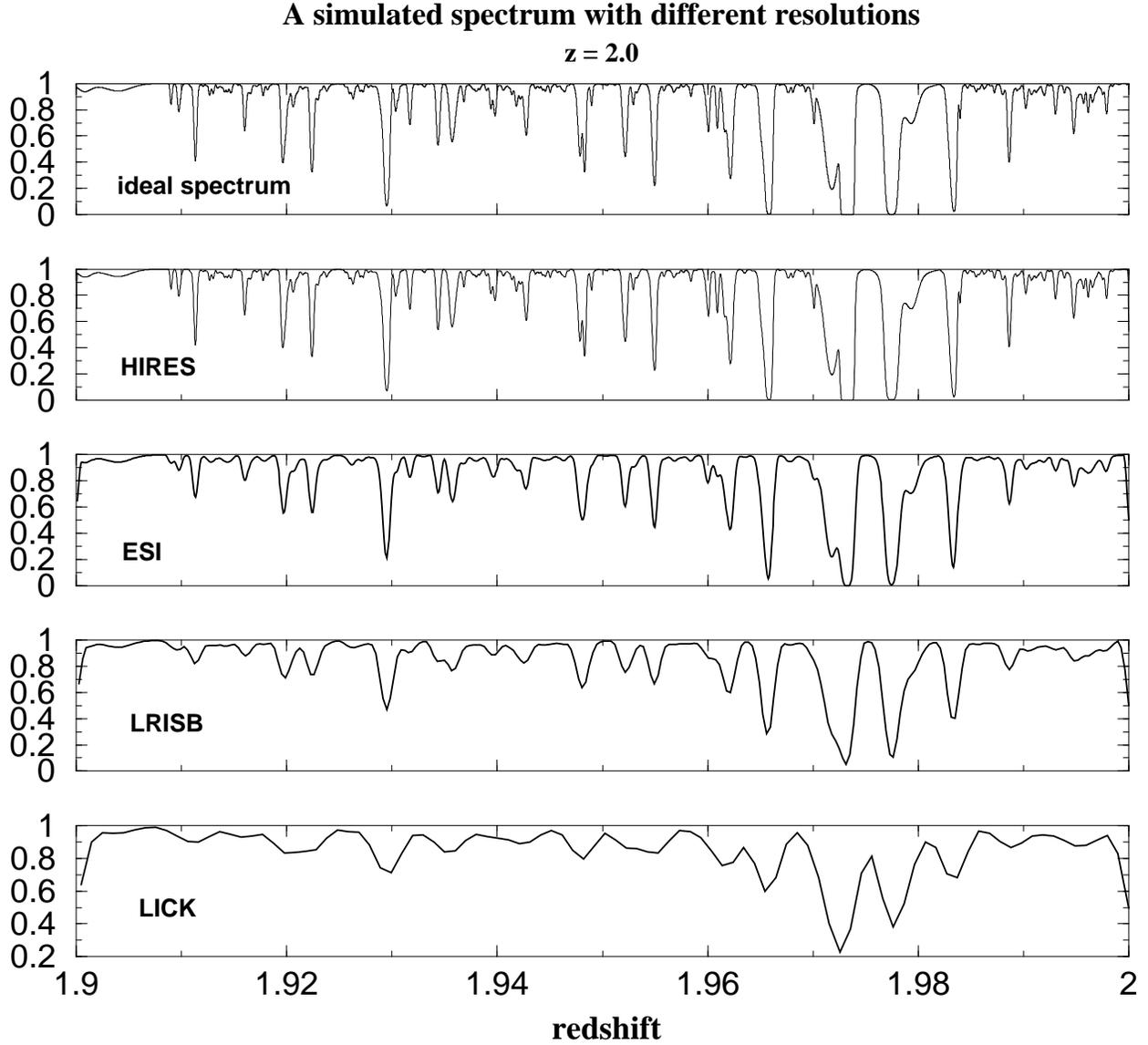}
\caption{We degraded an ideal flux spectrum from our simulation to different resolutions based on desired instruments. HIRES: High Resolution Echelle Spectrometer of the Keck telescope, FWHM = 8 km/s. ESI: An Echellette Spectrograph and Imager for the Keck II Telescope, FWHM = 55 km/s. LRIS: Low Resolution Imaging Spectrometer resolution of the Keck telescope, FWHM = 115 - 300 km/s (115 km/s in this Figure). Lick: FWHM: 250 km/s. }
\label{fig:SpeDegrade.fig}
\end{figure}

In order to compare with the  observational data, we degrade our ideal spectra to the resolution of the desired instrument. The resolution of each ideal QSO spectrum was degraded by convolving the entire spectrum with a normalized Gaussian function :
\begin{equation}
f(x) = \frac{1}{\sigma \sqrt{2 \pi}}  exp[\frac{-x^2}{2\sigma^2}]
\label{eqn:Degrade_Gaussian.eqn}
\end{equation}

with $\sigma$ = $\frac{FWHM}{\sqrt{8 ln 2}} $, where FWHM is the full width half maximum of the spectral resolution of the desired instrument. The convolution subroutine used was based on FFT algorithms from Numerical Recipes (Press et al.,1988). Figure (\ref{fig:SpeDegrade.fig}) shows the degraded simulated spectra at different resolutions.

\begin{figure}
\plotone{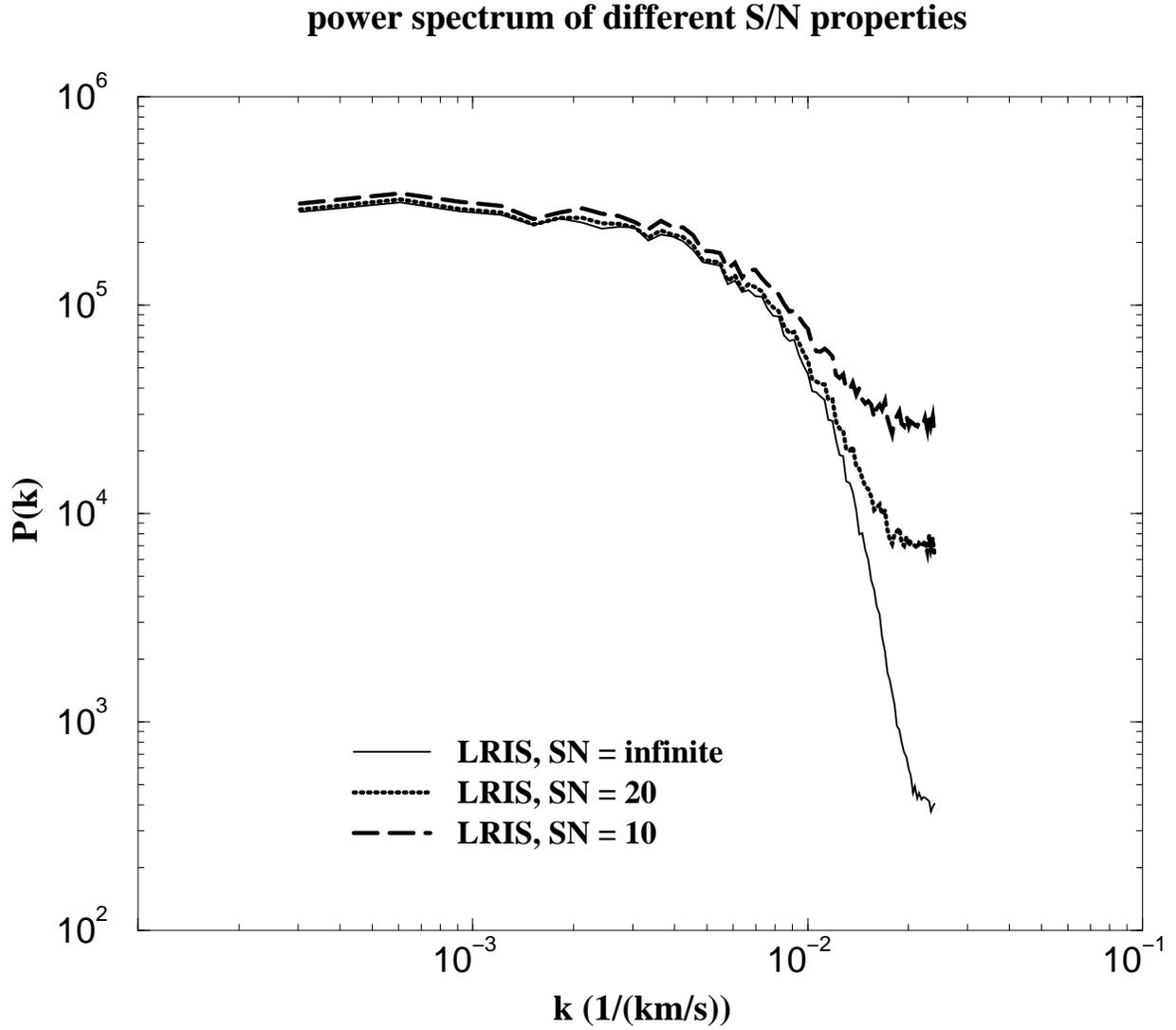}
\caption{The averaged flux power spectrum based on 300 simulated QSO spectra. We degraded the ideal spectra to LRIS resolution and then added different amounts of noise. The solid curve is the LRIS spectra without noise, the dotted curve is with S/N = 20 and the dashed curve is with S/N = 10. The normalization factor is not specified here.}
 
\label{fig:SN.fig}
\end{figure}

We also examined the effect of different S/N in our simulated spectra based on Gaussian-distributed noise. The noise comes from a Gaussian distribution with zero mean and unit variance:
\begin{equation}
p(y)dy = \frac{1}{\sqrt{2 \pi}} \exp[- y^2/2] dy
\end{equation}
 and we denote the noise as $N_{gauss}$. Then with a given S/N value, we derived the final flux spectra by adding the corresponding noise to the original flux spectra:
\begin{equation}
f_{final} = f_{orig} + \frac{N_{gauss}}{S/N} 
\end{equation}
where $f_{final}$ is the final flux with noise and $f_{orig}$ the flux after degrading the spectrum. Then we define the overflux $\delta_f$ as $\frac{f-\bar{f}}{\bar{f}}$, where $\bar{f}$ is the mean flux of the spectrum at a given redshift interval ( 1.754 $\le$ z $\le$ 1.954 in this paper) and is calculated by averaging the whole data points in the spectrum. So we have : 
$\delta_{f}$($z_1$), $\delta_{f}$($z_2$), $\ldots$ $\delta_{f}$($z_n$) from a QSO spectrum, where $z_{i}$, $i= 1,2$ $\ldots$, n are the corresponding redshifts at each data point. 
Then we calculate the corresponding flux power spectrum of these data points using the FFTW algorithm (Frigo et al. 1998).  Here k is defined as $\frac{2 \pi }{x (km/s)}$. Figure (\ref{fig:SN.fig}) shows the flux power spectrum. Adding noise increases the amplitude of the flux power spectrum, which means introducing small scale power. Beyond k = 0.01, the noise dominates the amplitude of the power spectrum. This tells us that the data beyond k = 0.01 is not reliable, due to the noise effect.

\begin{figure}
\plotone{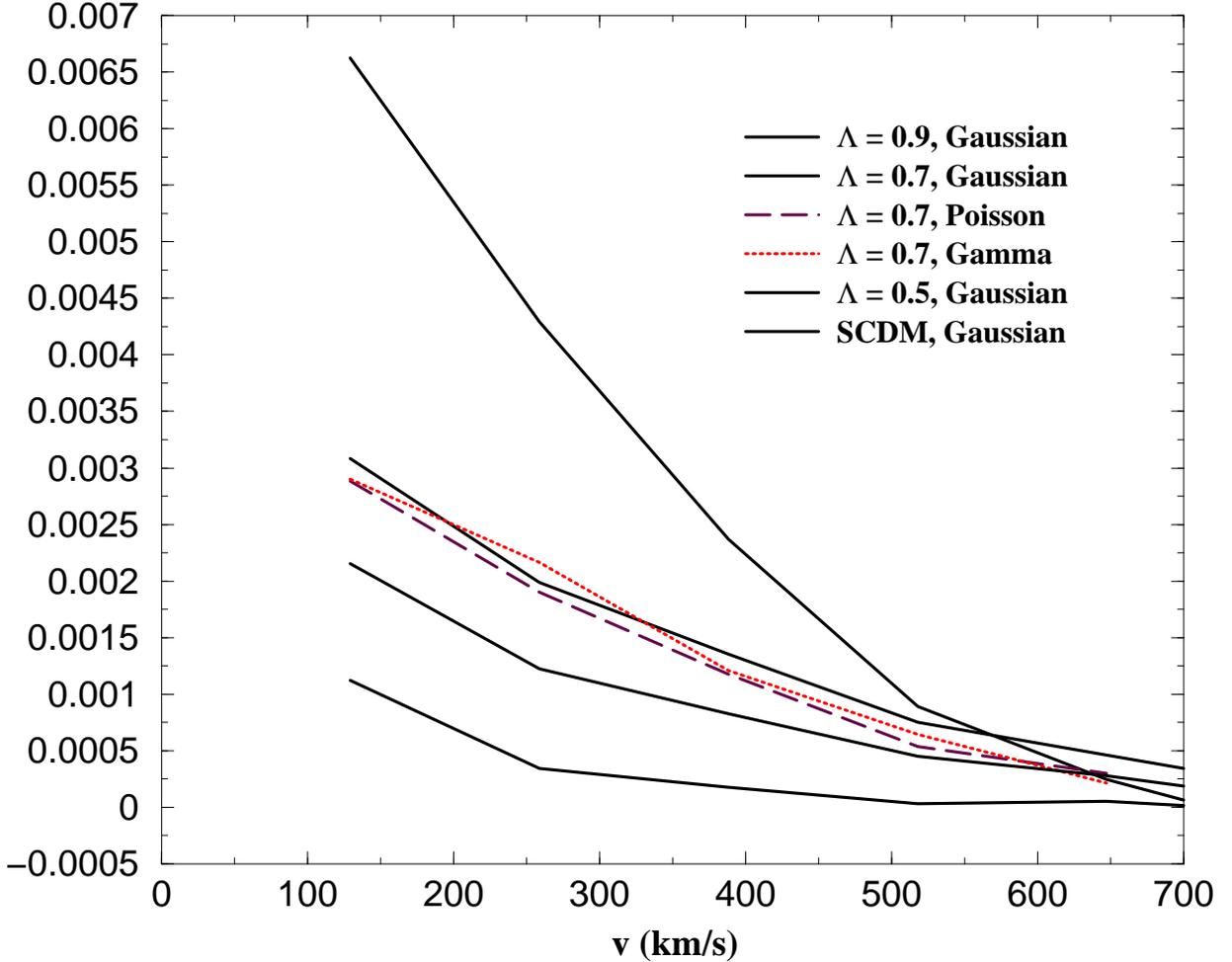}
\caption{Flux cross-correlation functions with different noise distribution.Each curve is the average of 300 cross-correlation functions while each cross-correlation function was derived from a paired QSO spectra based on our simulations. The corresponding FWHM is 300 km/s, the pixel size is 130 km/s and the signal-to-noise ratio is 10. The angular separation is 120". The four solid curves are SCDM (the lowest solid curve), $\Omega_\Lambda$ = 0.5 (the second lowest solid curve), $\Omega_\Lambda$ = 0.7 (the second highest solid curve) and $\Omega_\Lambda$ = 0.9 (highest solid curve) cosmological models with Gaussian noise. The long dashed curve is the $\Omega_\Lambda$ = 0.7 cosmological model with Poisson noise distribution and the dotted curve is the $\Omega_\Lambda$ = 0.7 cosmological model with Gamma noise distrubution.}
\label{fig:CrossCorre_Noise.fig}
\end{figure}

Because the distribution of noise in the observation is unknown, we chose several common distributions of noise to add to our simulated spectra for the analysis work. We compared the results based on Gaussian-distributed noise with other noise distributions including the Poisson and Gamma distributions. We used the routines in Numerical Recipes (Press et al., 1988) to generate noise from the above distributions. For a Poisson distribution, the total probability of integer j (event j) is : 
\begin{equation}
Prob(j) = \int_{j- \epsilon}^{j+ \epsilon} P_x (m) dm = \frac{x^j e^{-x}}{j !} 
\end{equation}
The Poisson noise, $N_{poisson}$ is a random deviate drawn from the Poisson distribution with unit mean. Similarly, a Gamma distribution of integer order a $\ge$ 0 is the waiting time to the $a_{th}$ event in a Poisson  random process of unit mean. We know that a Gamma deviate has a probability $P_{a}(x)dx$ of occurring with a value between $x$ and $x+dx$, where
\begin{equation}
P_{a}(x) dx = \frac{x^{a-1} e^{-x}}{\Gamma (a)} dx, x \ge 0
\end{equation}
where $\Gamma$(x) is the gamma function. Then the noise is a deviate distributed as a gamma distribution of integer order 1, which is the waiting time to the first event in a Poisson process of unit mean.  Figure (\ref{fig:CrossCorre_Noise.fig}) shows the flux cross-correlation functions at S/N = 10 with different noise distributions. We conclude that for a given S/N ratio, the distribution of noise does not have a significant influence on the final cross-correlation function.


\section{Methodology and Numerical Procedures}
\label{sec:Method}

\subsection{Cosmological Simulations and Simulated Spectra}
In this work, we used data from six simulations with different values of $\Omega_\Lambda$ as discussed in $\S$\ref{subsec:LyaSim} (Simulations Set A in Table (\ref{table:CosPara.tbl})). Before more QSO pairs are available from Sloan Digital Sky Survey (SDSS), approximately a few dozen QSO pairs with good quality should be observed through Keck (Kirkman et al. 2002). To provide an useful statistical study, we have to generate several groups of simulated pairs where each group has approximately the same number of pairs as are available observationally. We also know that pairs with angular separations between 1' - 3' contain good information about geometry in their cross-correlation functions (McDonald et al. 1999). Therefore, we generated 300 paired QSO spectra with an angular separation of 120" using the spectrum generator code described in $\S$\ref{subsec:SpeGen}. The length of a spectrum is of $\Delta$ z = 0.2, which is around 7 $\sim$ 8 times the box size. Since we shoot lights in random directions, each line of sight will not go through the same point in the box.

 Then we degraded the ideal QSO spectra to a FWHM of 300 km/s and a pixel size of 130 km/s, which is comparable to the LRIS of the Keck telescope. Gaussian-distributed noise with S/N = 10 and 20 were also added to the degraded simulated LRIS spectra for our analysis. These values were chosen comparable to the observational data which will be available in the near future.  
\subsection{Cross-Correlation Functions}
\label{subsec:CrossCorre}
Given a pair of QSO spectra, we first calculated the overflux at each point of a spectrum. The overflux $\delta_f$ and the mean flux $\bar{f}$ are defined in $\S$ \ref{subsec:SpeDeSN}. However instead of considering one QSO spectrum as we did in $\S$ \ref{subsec:SpeDeSN}, we now have a pair of QSO spectra, so the mean flux $\bar{f}$ is averaged over the whole data points of the paired spectra. Therefore we have : \\

$\delta_{f1}$($z_1$), $\delta_{f1}$($z_2$), $\ldots$ $\delta_{f1}$($z_n$) from the first QSO spectrum \\
$\delta_{f2}$($z_1$), $\delta_{f2}$($z_2$), $\ldots$ $\delta_{f2}$($z_n$) from the second QSO spectrum \\

where $z_{i}$, $i = 1, 2$, $\ldots$ n, are the corresponding redshifts at each point of the QSO spectra. Thus for a point $\delta_{f1}$($z_i$) from the first spectrum and another point $\delta_{f2}$($z_j$) from the second spectrum, their parallel velocity separation can be expressed as:
\begin{equation}
\Delta v_\parallel \equiv c\frac{\mid z_{j} - z_{i} \mid}{1+\bar{z}}
\label{eqn:Vp_def.eqn}
\end{equation}
where $\mid z_{j} - z_{i} \mid$ are small and $\bar{z}$ = $\frac{z_1 + z_n}{2}$.
The cross-correlation at a given parallel velocity separation $\Delta$$v_\parallel$ is denoted by $\xi$($\Delta$$v_{\parallel}$) and is calculated by : 
\begin{equation}
\xi(\Delta v_{\parallel}) = <\delta_{f1}(z_{i}) \delta_{f2}(z_{j})>
\label{eqn:Xi_def.eqn}
\end{equation}
 averaging over all possible permutations of (i,j) which satisfy Equation (\ref{eqn:Vp_def.eqn}).

The correlations we are concerned about here are velocity separations of 600 km/s or less, 
which is less than 10 Mpc and is definitely much smaller than $\frac{1}{2}$ box size.
Therefore, even though the lines of sight wrapped the box a few times, the range of the correlation 
function we are interested in is small enough and should not be affected by our box size. 

Figure (\ref{fig:CrossCorre_2SNrandom.fig}) shows the cross-correlation functions of two QSO pairs from the $\Omega_\Lambda$ = 0.7 simulation. The large difference of the cross-correlation functions between the two pairs reveals the significant impact of cosmic variance, which is the reason why we introduced the MLE method (see Appendix (\ref{sec:MLE}) and $\S$\ref{subsec:Likelihood_Cal} for more details).

\subsection{Probability Density Functions}
\label{subsec:CreatPDF}
In order to apply the maximum likelihood method (MLE) to our analysis work, we need to provide accurate probability density functions first (Appendix (\ref{sec:MLE})).  For each of the 6 simulations, we have generated 300 paired QSO spectra with an angular separation of 120", providing 300 cross-correlation functions per simulation: $\xi$($\Delta$ $v_{\parallel}$ ; $\Delta$ $\theta$ $\mid$ $\Omega_\Lambda$), where $\Delta$ $\theta$ = 120" and $\Omega_\Lambda$ = 0.0, 0.5, 0.6, 0.7, 0.8 and 0.9. We denote $\xi$($\Delta$ $v_{\parallel}$ ; $\Delta$ $\theta$ $\mid$ $\Omega_\Lambda$) for given $\Delta \theta$ and $\Omega_\Lambda$ at a fixed $\Delta$ $v_{\parallel}^{\star}$ as $\xi(\Delta v_{\parallel}^{\star})$. At a given $\Delta$ $v_{\parallel}^{\star}$ each cross-correlation function provides one value of $\xi(\Delta v_{\parallel}^{\star})$, so we have 300 data points of $\xi(\Delta v_{\parallel}^{\star})$ at $\Delta v_\parallel^{\star}$ under given $\Omega_\Lambda$ and  $\Delta$ $\theta$. We use these data to construct a probability density function  of $\xi(\Delta v_{\parallel}^{\star})$ at $\Delta v_{\parallel}^{\star}$, denoted by PDF($\xi(\Delta v_{\parallel}^{\star})$), for known $\Omega_\Lambda$ and $\Delta$ $\theta$. We repeat the procedures for different $\Delta$ $v_{\parallel}$ and $\Omega_\Lambda$. Figures (\ref{fig:CrossCorre_v129.fig}), (\ref{fig:CrossCorre_v388.fig}) and (\ref{fig:CrossCorre_v647.fig}) show the PDF($\xi(\Delta v_{\parallel}^{\star})$) at three different $\Delta v_\parallel^{\star}$ for different cosmological models.

\subsection{The Likelihood Calculation}
\label{subsec:Likelihood_Cal}

 Given a QSO pair, pair A, we first calculate the corresponding cross-correlation function $\xi_{A}$($\Delta v_\parallel$) based on Equation (\ref{eqn:Xi_def.eqn}). We denote the points we use in this work as ($\Delta v_\parallel^{1}$, $\xi_{A}$($\Delta v_\parallel^{1}$)), ($\Delta v_\parallel^{2}$, $\xi_{A}$($\Delta v_\parallel^{2}$)), ...... ($\Delta v_\parallel^{m}$, $\xi_{A}$($\Delta v_\parallel^{m}$)). In our work, we use five points: ($\Delta v_\parallel^{1}$ = 129 km/s, $\xi_{A}$($\Delta v_\parallel^{1}$)) ,  ($\Delta v_\parallel^{2}$ = 259 km/s, $\xi_{A}$($\Delta v_\parallel^{2}$)) , ($\Delta v_\parallel^{3}$ = 388 km/s, $\xi_{A}$($\Delta v_\parallel^{3}$)), ($\Delta v_\parallel^{4}$ = 518 km/s, $\xi_{A}$($\Delta v_\parallel^{4}$)) and ($\Delta v_\parallel^{5}$ = 647 km/s, $\xi_{A}$($\Delta v_\parallel^{5}$)).  For each given $\Omega_\Lambda$, using the PDF($\Delta v_{\parallel}^{i}$ $\mid$ $\Omega_\Lambda$) generated in $\S$\ref{subsec:CreatPDF}, we can find that the probability of getting ($\Delta v_\parallel^{i}$, $\xi_{A}$($\Delta v_\parallel^{i}$)) is $f_{A}^{i}$.

 Assuming we have more than one paired QSO spectra: $P_{A1}$, $P_{A2}$, ...., $P_{Ar}$, we can combine the probability $f_{Aj}^{i}(v_{i})$, j = 1, $\ldots$, r, where $v_{i}$ is a given velocity bin for all pairs. Then $L_{v_i}$ = $\Pi_{j=1}^{5}$$f_{Aj}^{i}$, where all ($v_{j}^{i}$, $f_{Aj}^{i}$) are independent to each other since $P_{A1}$, $P_{A2}$, ...., $P_{Ar}$ are from different parts of the sky. By repeating the above calculation in different cosmological models, we can derive a likelihood function $L_{v_i}(\Omega)$ at the velocity bin $v_{i}$. Since the data points at different velocity bins do not share the same probability denstity function, we can combine these data and derive a final likelihood function :
$L(\Omega)$ = $\Pi_{i=1}^{n} L_{v_{i}}(\Omega)$

   Figure (\ref{fig:MLK_Gp1_SN10_allv.fig}) shows the likelihood function of a group of 30 pairs at different velocity separations. The likelihood functions have similar results at different velocity separations and the combination at all velocity separations provides a better statistics and an useful result. By maximizing the likelihood function, we can determine $\Omega_{\Lambda}$. The advantage of using the MLE method here is that we can derive $\Omega_\Lambda$ without collecting many paired QSO spectra with the same angular separation and redshift to determine a reliable cross-correlation function beforehand.

The above analysis ignores the correlations between data points at different velocity separations within a given pair spectrum. The fact that the combined
liklihood function (Fig. 11) is very close to the v=129 km/s liklihood function, and that the other velocity separations have similar profiles, suggests
that this may indeed be the case. We would like to make two points in defense of our method, however. We could just as well 
base the determination of $\Omega_\Lambda$ on a single velocity separation, in which case the issue goes away and our MLE method is essentially unchanged. 
Second, we view combining the different velocity bins as a way of making use of all the data that is available. One can look for the bias introduced by
treating them as independent {\em a posteriori}. In Sec. 5.3 we apply our method to a blind sample of pair spectra taken from a $\Lambda=0.7$ simulation. 
As shown in Fig. 16 and discussed in Sec. 5.3, the bias is dominated by sample size. We find that a sample of 30 pairs with S/N=20 is sufficient to 
recover the correct value of $\Lambda$ within uncertainties. 


\section{Results}
\label{sec:Res}

\subsection{Sensitivities to other cosmological parameters}
\label{subsec:SenCos}
Our preliminary test of several other cosmological parameters includes $\sigma_{8}$, photoionization parameter $J$, $\Omega_{B}$ and h. There is a large scatter in the value of $\sigma_{8}$ measured by various scientists (Ballinger et al. 1996; Ratra et al. 1997; Efstathiou et al. 2001; Holder et al. 2001; Wang et al. 2002) and we thereby choose the value of $\sigma_{8}$ to be 0.73 with a 10$\%$ varying range. Figure (\ref{fig:CrossCorre_s8com.fig}) shows that fluctuations in cross-correlation functions caused by the uncertainty of $\sigma_{8}$ is relatively small and ignorable compared with that of $\Omega_\Lambda$. The ranges of $\Omega_{B}$ and h are first constrained by the relation $\Omega_{B} h^{2}$ $\sim$ 0.019 (Burles et al. 1999; Tytler 1999). At the same time, we pick the value of Hubble Constant with h = 0.72 $\pm$ 0.08 based on the HST key project (Freedman et al. 2001). Our result indicates that the uncertainties in $\Omega_{B}$ and h do not have significant influence in this work (Figure (\ref{fig:CrossCorre_Bcom.fig})).

Another important parameter $J$ indicates the intensity of the photoionization background. The standard $J$ = 1.0 in our simulations corresponds to the Haardt $\&$ Madau UV radiation field (1996). This ionization background has been implemented in many previous Ly$\alpha$ simulations (eg. Bryan et al. 1999). The uncertainty in $J$ is large and in order to be safe, we vary the value of $J$ from 0.25 to 3.0. From Figure (\ref{fig:CrossCorre_Jcom.fig}), we know that the fluctuation in the final cross-correlation function caused by $J$ is approximately at the same order as caused by $\Omega_\Lambda$. Therefore, quantifying the value of $J$ becomes important and a detailed study will be done in our next paper.

We also compare our results with the parameter studies discussed in McDonald (2003). In McDonald (2003), he varies five free parameters in the Ly$\alpha$ forest model including the power spectrum amplitude $A_{1}$, the power-law index of the power spectrum at $k_{1}$, the factors of a power-law temperature-density relations $T_{1.4}$ and $\gamma$-1, and the mean flux $\bar{F}$ (See McDonald 2003 for detailed definitions of these parameters). While in our work, we focus on changing the cosmological parameters including $\sigma_{8}$, $\Omega_{B}$(h) and the photoionization parameter $J$ to study the corresponding results of the output simulated spectra calculated by the Spectrum Generator mentioned in $\S$\ref{subsec:SpeGen}. The two groups approach this issue of parameter sensitivities in different ways and both of them show that measuring other parameters accurately is one of the most important issues in a true AP test.

\subsection{Probability density functions with S/N properties}
\label{subsec:PDF_SN}

It is important to understand the effect of noise in our analysis work. We first pay attention to the probability density functions drawn from the data sets with S/N = 10 and S/N = 20 in Figure (\ref{fig:PDF_SNcom.fig}). This figure shows that the probability density function with a high S/N ratio has a smaller standard deviation than that of a lower S/N.  This indicates the noise spreads out the probability density function of correlations and we should provide different observational data points their corresponding probability density function during the analysis procedures.

\begin{figure}
\plotone{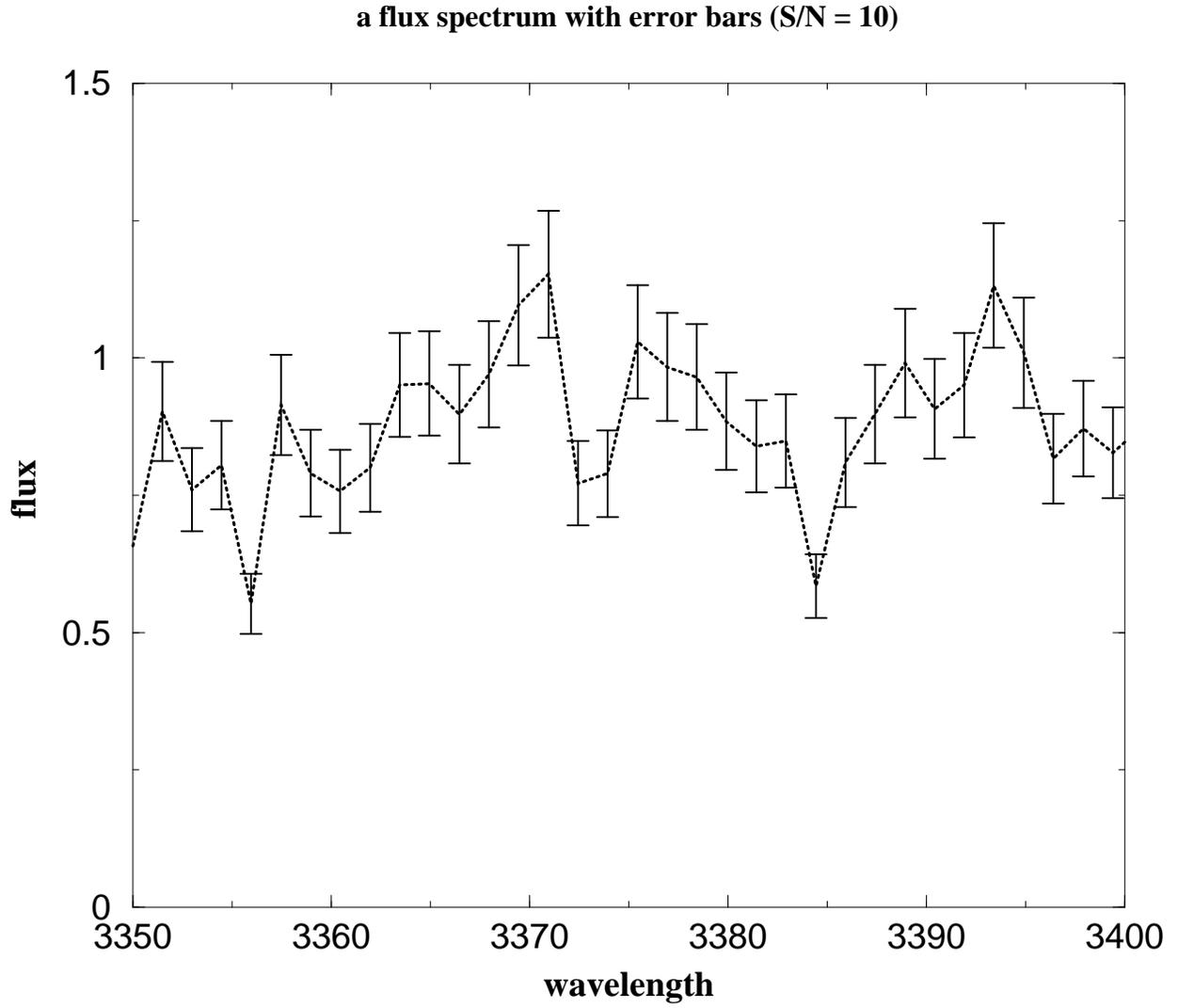}
\caption{A flux spectrum with error bars. The error bars are results of instrumental noise. The S/N is 10.}
\label{fig:FluxSpecErr_1.eps}
\end{figure}

\begin{figure}
\plotone{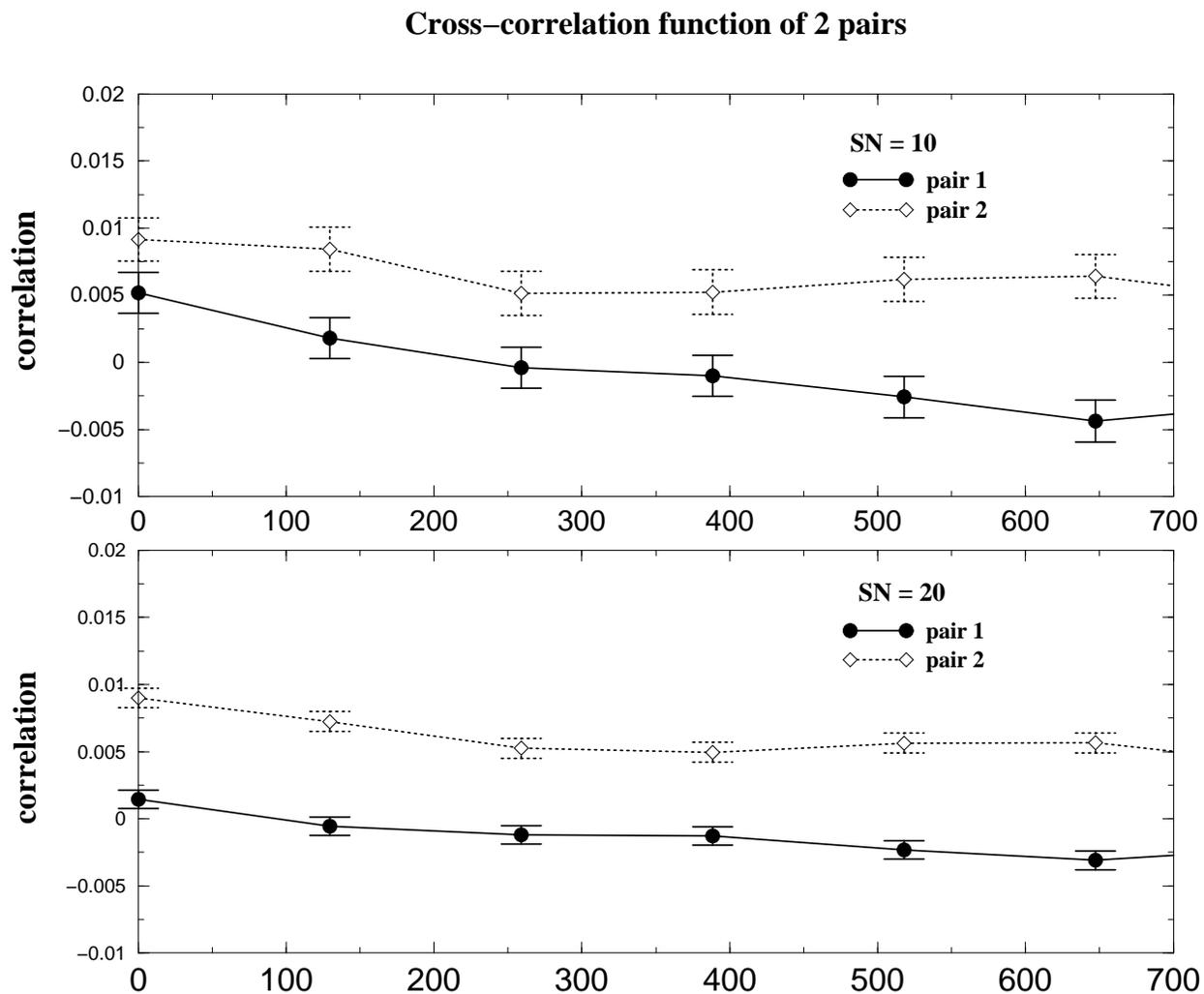}
\caption{Flux cross-correlation functions of two QSO pairs from the $\Omega_\Lambda$ = 0.7 simulation. The solid curve denotes the result of the first pair while the dashed curve denotes the result of the second pair. The error bars were derived from the flux spectra with S/N = 10 and S/N = 20. The angular separation is 120".}
\label{fig:CrossCorre_2SNrandom.fig} 
\end{figure}

\begin{figure}
\plotone{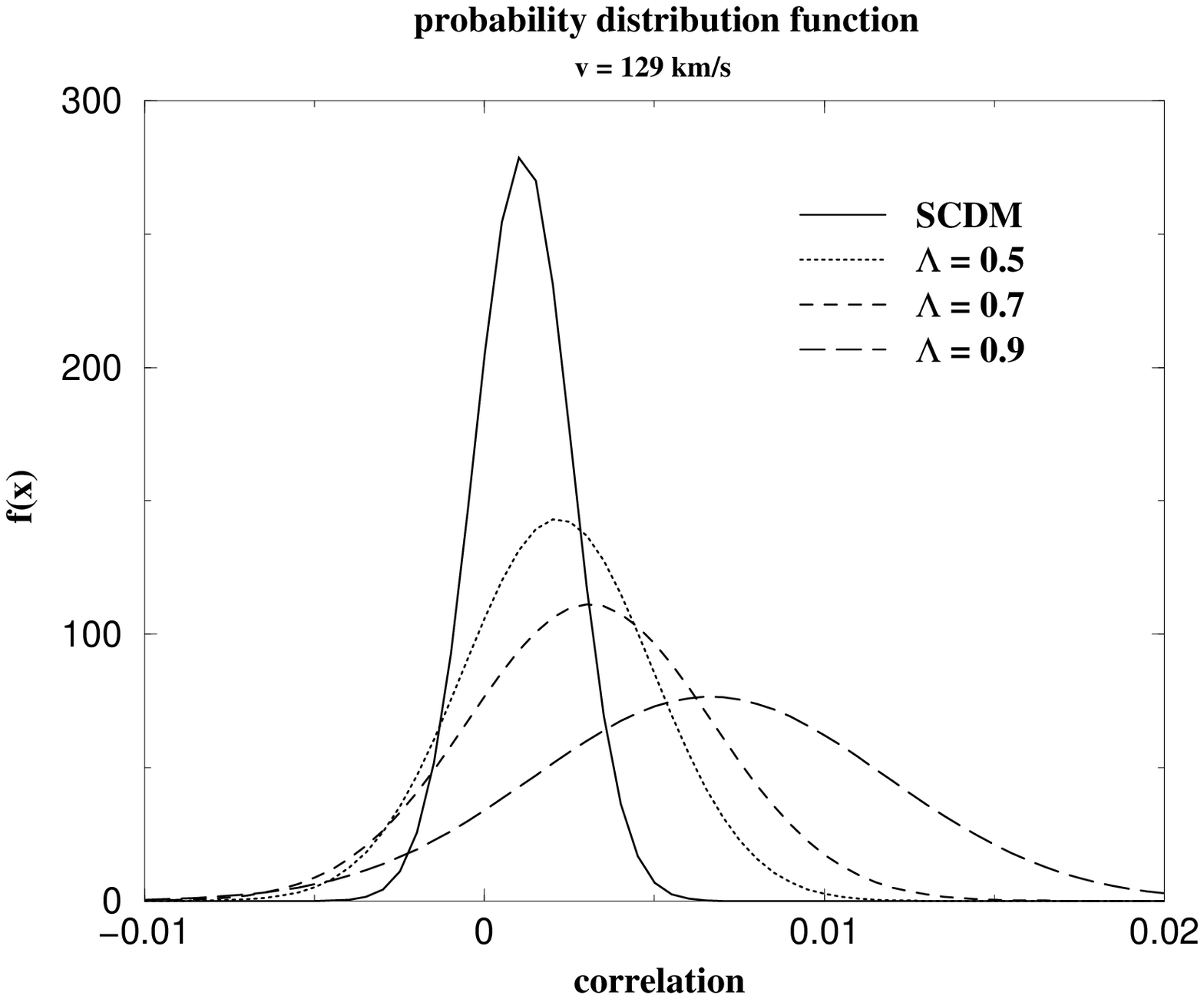}
\caption{PDF($\xi(\Delta v_{\parallel}^{\star})$) of cosmological models with different $\Omega_\Lambda$. The x-axis is the flux cross-correlation and the y-axis f(x) is the probability density function for x. The $\Delta v_{\parallel}^{\star}$ equals to 129 km/s. The pair separation is 120". The solid curve is the SCDM model, the dotted curve is the $\Omega_\Lambda$ = 0.5 model, the short-dashed curve is the $\Omega_\Lambda$ = 0.7 model and the long-dashed curve is the $\Omega_\Lambda$ = 0.9 model. We have fitted and normalized the original distribution curve to a normalized Gaussian distribution function in each cosmological model.} 
\label{fig:CrossCorre_v129.fig}
\end{figure}

\begin{figure}
\plotone{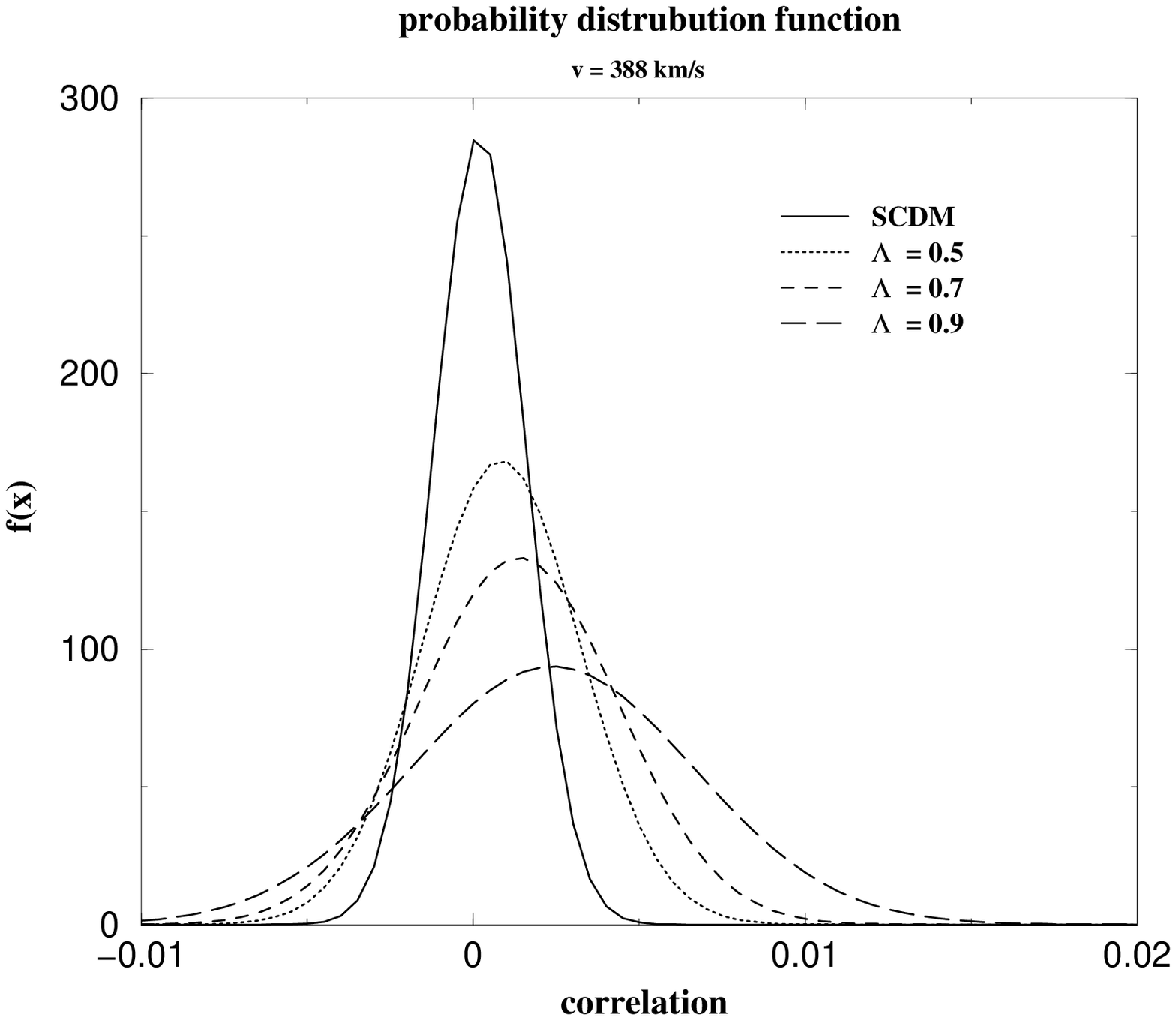}
\caption{PDF($\xi(\Delta v_{\parallel}^{\star})$) of cosmological models with different $\Omega_\Lambda$. The x-axis is the flux cross-correlation and the y-axis f(x) is the probability density function for x. The $\Delta v_{\parallel}^{\star}$ equals to 388 km/s. The pair separation is 120". The solid curve is the SCDM model, the dotted curve is the $\Omega_\Lambda$ = 0.5 model, the short-dashed curve is the $\Omega_\Lambda$ = 0.7 model and the long-dashed curve is the $\Omega_\Lambda$ = 0.9 model. We have fitted and normalized the original distribution curve to a normalized Gaussian distribution function in each cosmological model.} 
\label{fig:CrossCorre_v388.fig}
\end{figure}

\begin{figure}
\plotone{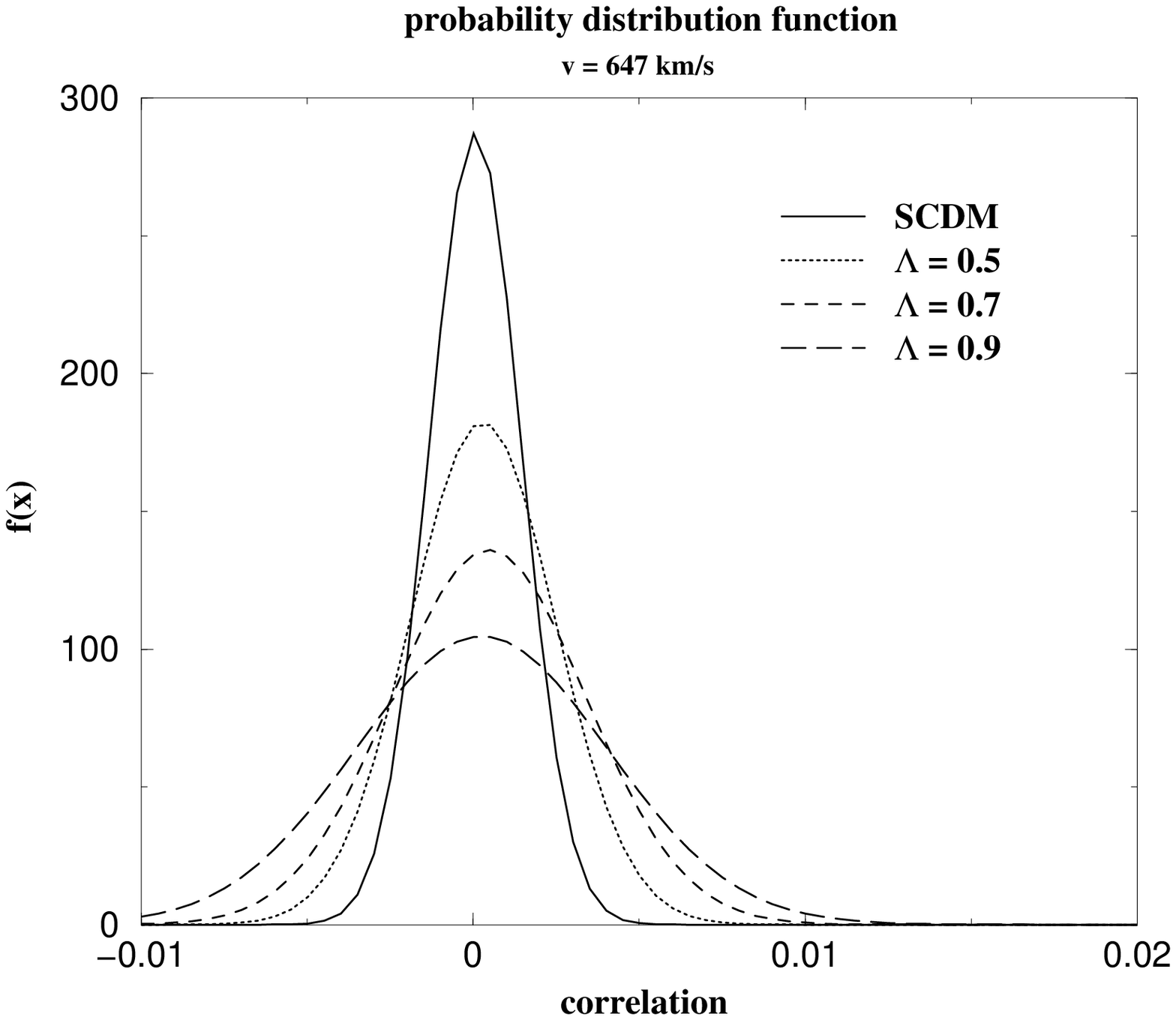}
\caption{PDF($\xi(\Delta v_{\parallel}^{\star})$) of cosmological models with different $\Omega_\Lambda$. The x-axis is the flux cross-correlation and the y-axis f(x) is the probability density function for x. The $\Delta v_{\parallel}^{\star}$ equals to 647 km/s. The pair separation is 120". The solid curve is the SCDM model, the dotted curve is the $\Omega_\Lambda$ = 0.5 model, the short-dashed curve is the $\Omega_\Lambda$ = 0.7 model and the long-dashed curve is the $\Omega_\Lambda$ = 0.9 model. We have fitted and normalized the original distribution curve to a normalized Gaussian distribution function in each cosmological model.}
\label{fig:CrossCorre_v647.fig}
\end{figure}

\begin{figure}
\plotone{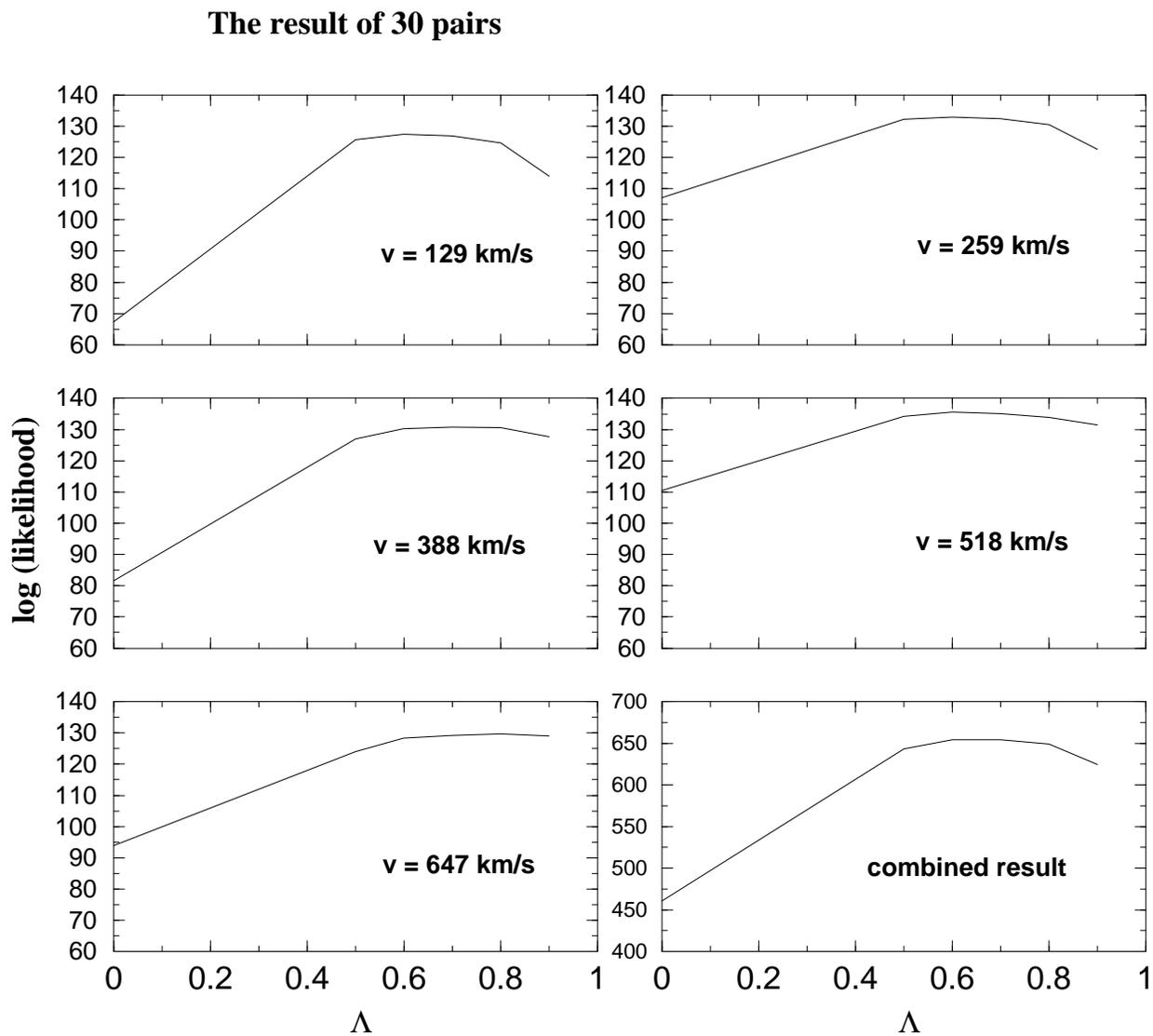}
\caption{The log(likelihood) functions at different velocity separations based on 30 QSO pairs. The functions at different velocity separations reveal similar results and the final combined result gives a better statistics.}
\label{fig:MLK_Gp1_SN10_allv.fig}
\end{figure}

\begin{figure}
\plotone{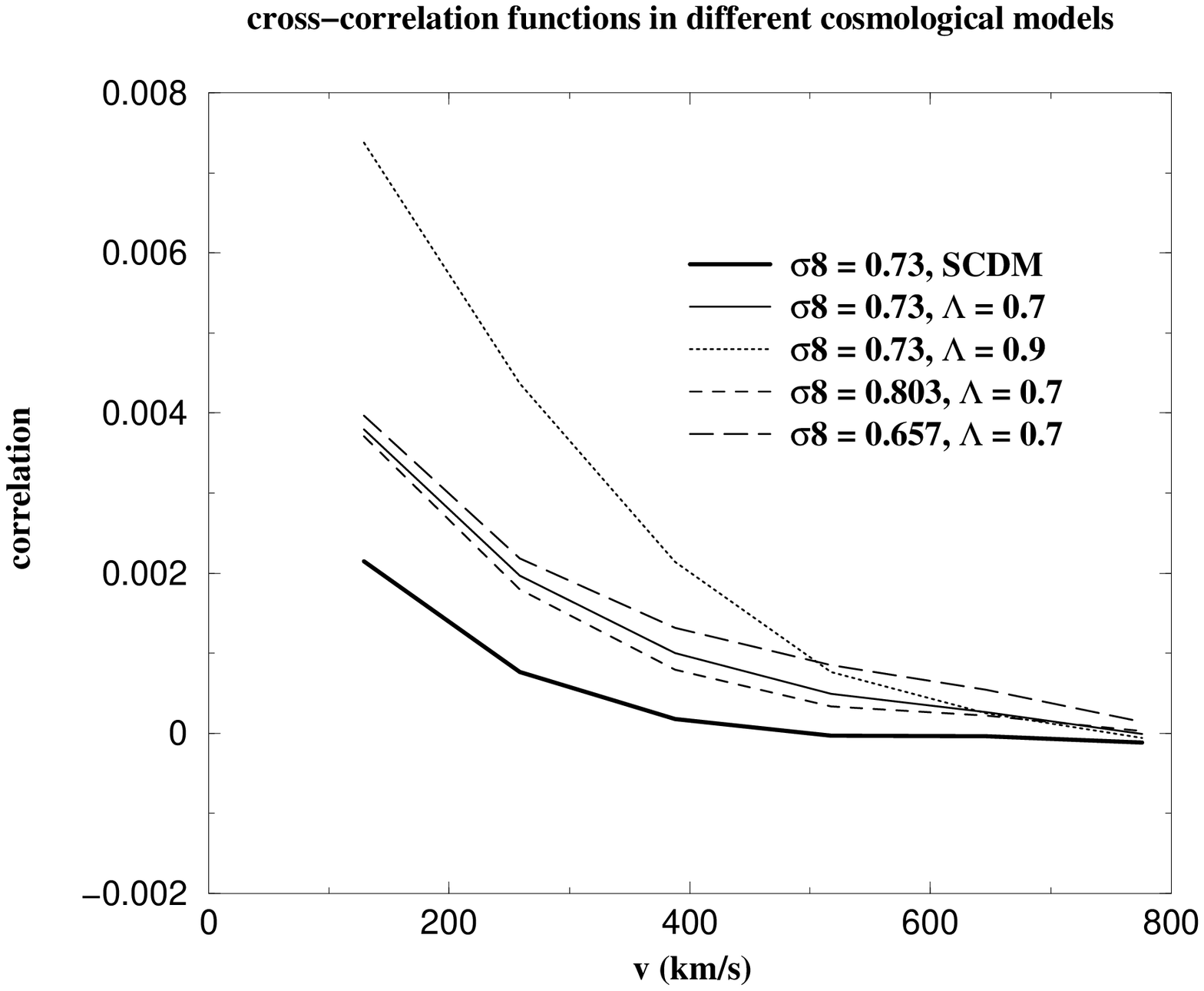}
\caption{We compare fluctuations in cross-correlation functions caused by the uncertainties of $\sigma_{8}$ and $\Omega_\Lambda$. Some other important parameters of these five simulations are $\Omega_{B}$ = 0.04, h = 0.69 and $J$ = 1.0.}
\label{fig:CrossCorre_s8com.fig}
\end{figure}

\begin{figure}
\plotone{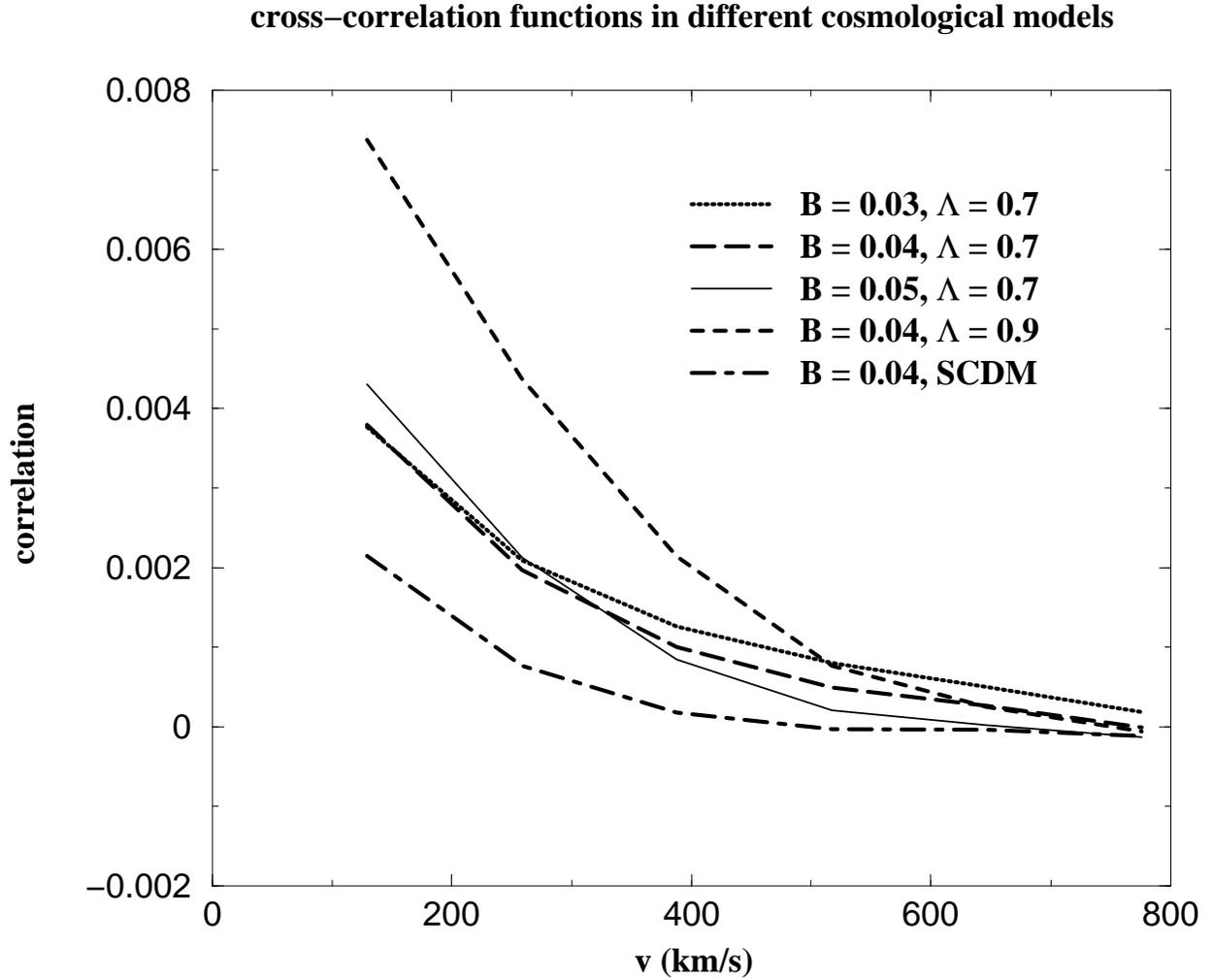}
\caption{We compare fluctuations in cross-correlation functions caused by the uncertainties of $\Omega_{B}$ and $\Omega_\Lambda$. Some other important parameters of these five simulations are $\sigma_{8}$ = 0.73 and $J$ = 1.0. When $\Omega_{B}$ = 0.03, h=0.7967; when $\Omega_{B}$ = 0.05, h = 0.617.}
\label{fig:CrossCorre_Bcom.fig}
\end{figure}

\begin{figure}
\plotone{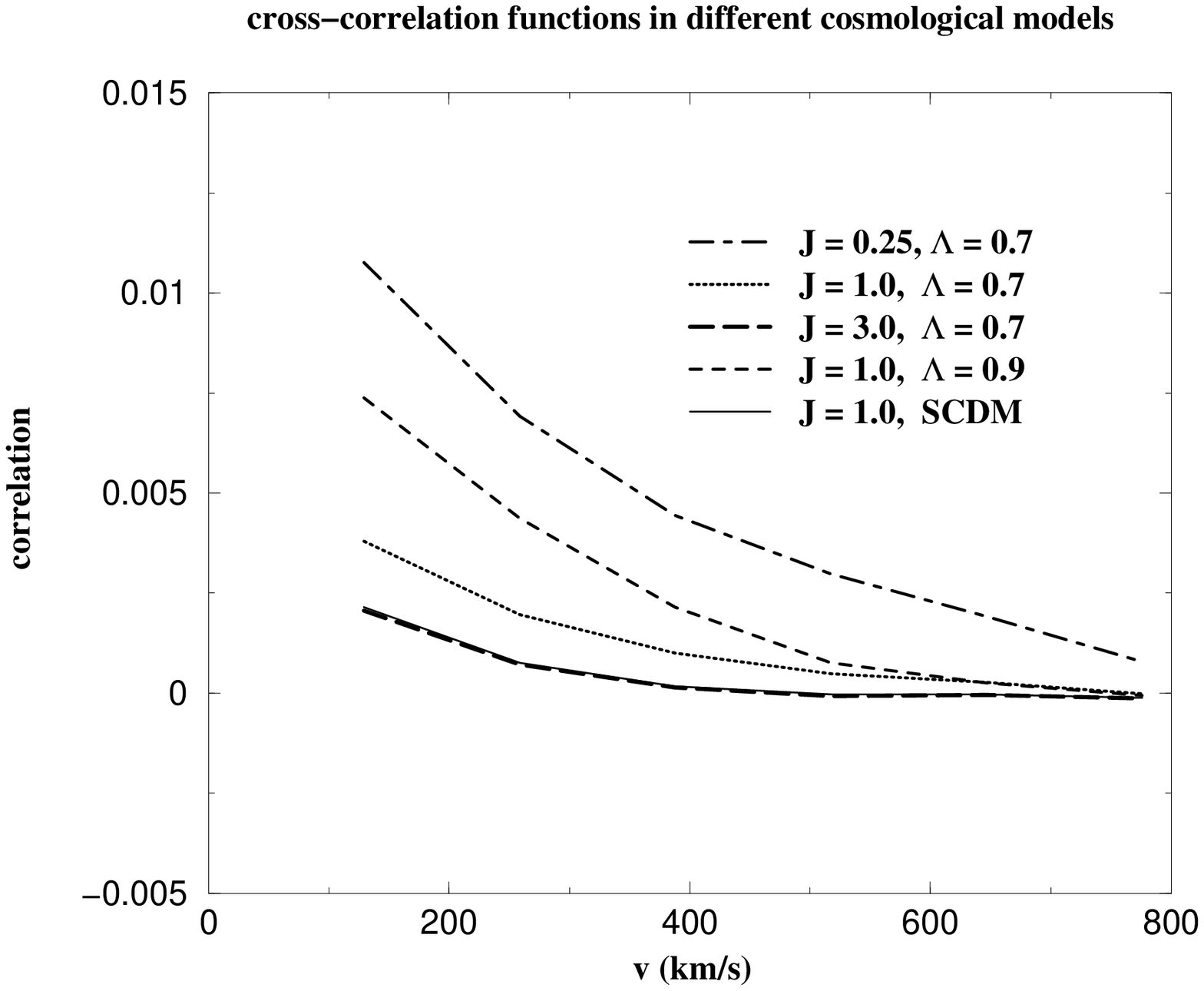}
\caption{We compare fluctuations in cross-correlation functions caused by the uncertainties of $J$ and $\Omega_\Lambda$. Some other important parameters of these five simulations are $\Omega_{B}$ = 0.04, h = 0.69 and $\sigma_{8}$ =0.73 .}
\label{fig:CrossCorre_Jcom.fig}
\end{figure}

\begin{figure}
\plotone{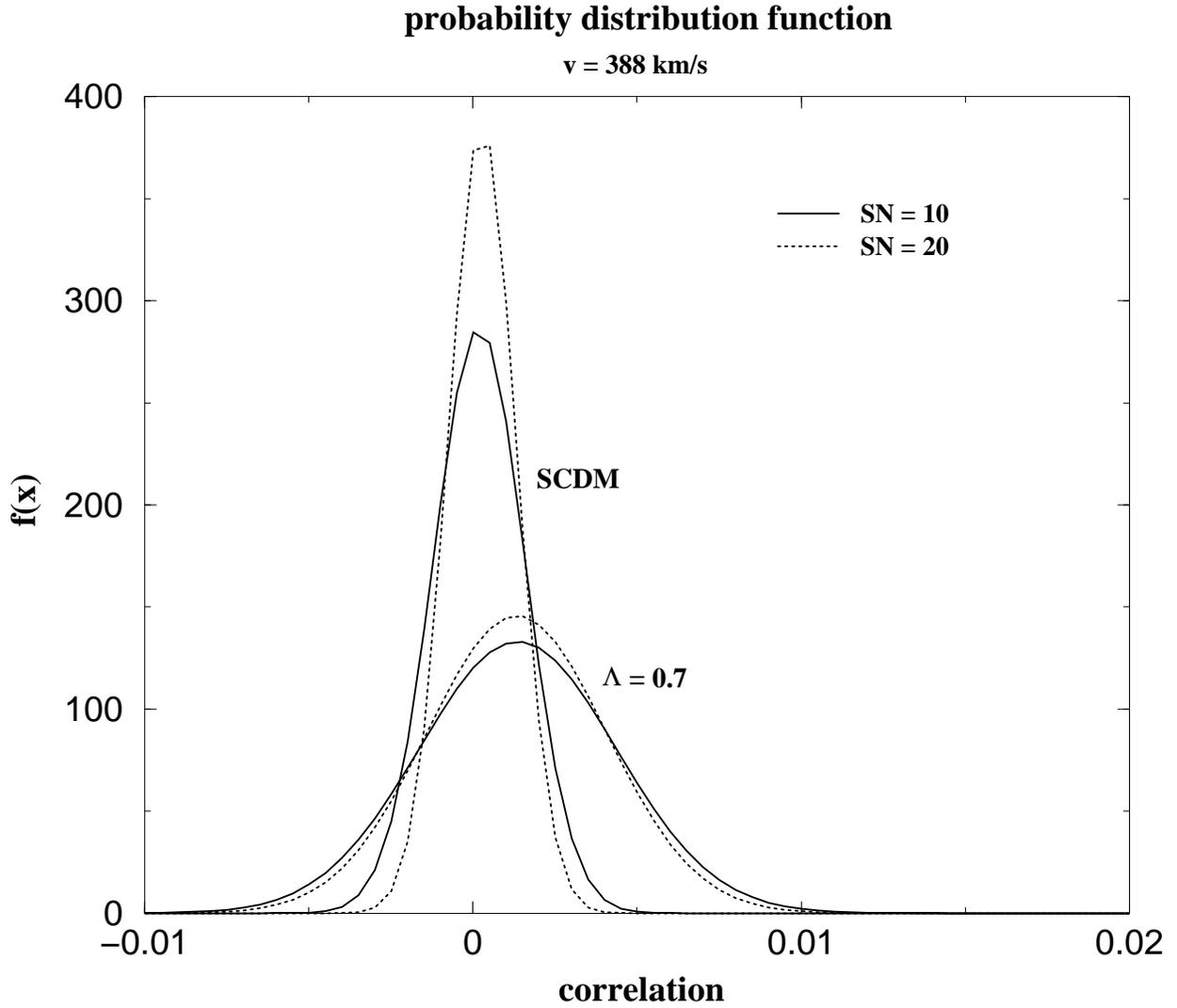}
\caption{The probability density functions at v = 388 km/s with SN = 10 and S/N = 20 for different cosmological models. The x-axis denotes the values of cross-correlation and the y-axis represents the probability density function f(x) for x. The PDF of a lower S/N has larger standard deviation than that of a higher S/N.}
\label{fig:PDF_SNcom.fig}
\end{figure}

The instrumental noise gives an error bar at each data point in a flux spectrum, and the noise at each data point is proportional to the value of the flux. Figure (\ref{fig:FluxSpecErr_1.eps}) includes the 1 $\sigma$ error bar at each data point for a flux spectrum. The uncertainties in the flux spectrum definitely result in the uncertainties in the cross-correlation fluction. Based on the standard error propagation formula (Appendix \ref{subsec:ErrPro}),  we can calculate the corresponding error bars in the cross-correlation function as shown in Figure (\ref{fig:CrossCorre_2SNrandom.fig}) for two randomly chosen pairs. In the upper panel of Figure (\ref{fig:CrossCorre_2SNrandom.fig}), the error bars are derived from the flux spectra with S/N = 10 while in the lower panel the error bars are calculated based on the S/N = 20 flux spectra.

  We should notice that the large differences in the cross-correlation functions of the two pairs in both panels are the result of cosmic variance. In the upper panel, the difference of cross-correlation in the two pairs is around the same order but slightly larger than the error bars in the cross-correlation drawn from the SN =10 flux spectra. Therefore, for each data point in a cross-correlation function the uncertainty due to the noise at each point could possiblely yield an uncertainty in the final likelihood function and thereby $\Omega_\Lambda$. Due to this, quantifying the uncertainty in our estimation of $\Omega_\Lambda$ due to the nosie is an important problem. More detailed work will be done in the near future.

 This is because one can never claim the exact value of the correlation while an error bar is concerned. Therefore, to quantify the uncertainty in the $\Omega_\Lambda$ due to the noise is an important problem and more detailed work will be done in the near future.


\subsection{Probability of Obtaining the Correct value of $\Omega_\Lambda$}
\label{subsec:Prob_Lamb}
To investigate the question about the number of pairs needed for deriving a reliable value of $\Omega_\Lambda$, we take 300 paired QSO spectra from our $\Omega_\Lambda$ = 0.7 simulation and pretend that they are observational data from the real Universe. First, we resample our 300 paired QSO spectra into 30 subgroups of 10 pairs each. As described in $\S$\ref{subsec:Likelihood_Cal}, in each subgroup we calculated the likelihood based on the data points of the 10 cross-correlation functions and then derived the combined-likelihood of the total 10 pairs in different cosmological models. Thus, in each subgroup, we derive a likelihood function of $\Omega_\Lambda$. By maximizing the likelihood function (or the $log(likelihood)$ function (LLK)), we derive the corresponding $\Omega_\Lambda$ (see Appendix (\ref{sec:MLE}) for details). This gives a distribution for values of $\Omega_\Lambda$ based on the 30 subgroups as shown in Figure (\ref{fig:Lambda_distri.fig}).  If we parametrize the distribution as a Gaussian, we find that $\Omega_\Lambda$ is 0.6 $\pm$ 0.145, where 0.6 is the mean and 0.145 is the standard deviation of the distribution from the 30 subgroups.


 Before any further analysis, we want to define the uncertainty in $\Omega_\Lambda$ first. In this paper, we denote an x $\%$ uncertainty in $\Omega_\Lambda$ as $\epsilon_{x \%}$ and it means the derived $\Omega_\Lambda^{\prime}$ is between $(\frac{100-x}{100})\Omega_\Lambda$ and $(\frac{100+x}{100})\Omega_\Lambda$. Here we have assumed that the real value of $\Omega_\Lambda$ in the Universe is 0.7 and thus the 20 $\%$ uncertainty of $\Omega_\Lambda$ ranges from 0.56 to 0.84 (within $\epsilon_{20 \%}$). Therefore, by calculating

\begin{equation}
P = \int_{0.56}^{0.84} dx \frac{1}{\sqrt{2 \pi \sigma}} \exp[-\frac{1}{2} \frac{(x-\mu)^2}{\sigma^2}]
\label{eqn:Prob_Lambda.eqn}
\end{equation}

where $\mu$ = 0.6 and $\sigma$ = 0.145, we find that the probability of getting $\Omega_\Lambda$ within $\epsilon_{20 \%}$ is 56 $\%$. Similarly, the $\epsilon_{10 \%}$ for the value of $\Omega_\Lambda$ ranges from 0.63 to 0.77. So by replacing the lower and the upper integration limits in Equation (\ref{eqn:Prob_Lambda.eqn}) to 0.63 and 0.77, we derived the probability of obtaining $\Omega_\Lambda$ within $\epsilon_{10 \%}$ is 30 $\%$. By repeating the above procedures, we also conclude that the probability of getting $\Omega_\Lambda$ within $\epsilon_{5 \%}$ ( 0.665 $\le$ $\Omega_\Lambda$ $\le$ 0.735) is 15 $\%$. 

To investigate the importance of sample size, we also split the 300 paired-spectra into 20 subgroups with 15 pairs in each subgroup, 15 subgroups with 20 pairs in each subgroup, 12 subgroups with 25 pairs in each subgroup and 10 subgroups with 30 pairs in each subgroup. Then in each case we calculated the distribution of $\Omega_\Lambda$ and the probability of deriving $\Omega_\Lambda$ with an uncertainty of $\epsilon_{20 \%}$, $\epsilon_{10 \%}$ and $\epsilon_{5 \%}$. For the case we have 30 paired QSO spectra in each subgroup (10 subgroups), the probability to confine $\Omega_\Lambda$ within $\epsilon_{20 \%}$ is 94$\%$, within $\epsilon_{10 \%}$ is 66$\%$ and within $\epsilon_{5 \%}$ is 36$\%$. A summary of results is displayed in Table (\ref{table:Lambda_Prob.tbl}).  Each row in the table indicates one statistical study. The first column is the number of QSO pairs in one subgroup while the second column is the number of subgroups in each study. The third column denotes the result of $\Omega_\Lambda$. Finally, the fourth, fifth, and the sixth columns are the probability of getting $\Omega_\Lambda$ within an uncertainty range of $\epsilon_{20 \%}$, $\epsilon_{10 \%}$ and $\epsilon_{5 \%}$.

\section{Discussion and Conclusion}
\label{sec:DisAndCon}
 Even though the possibility of using the AP test on the Ly$\alpha$ forest to measure $\Omega_\Lambda$ has been emphasized (Hui et al. 1999; McDonald et al. 1999), the practical work of dealing with real observational data is not easy. There are several major difficulties in this work including cosmic variance, uncertainty of other cosmological parameters, understanding of spectral resolution along with S/N properties and the continuum fitting. The main focus of this paper is to produce noisy low resolution spectra which are comparable to the real LRIS observational data and show how to overcome the errors due to noise and cosmic variance. Aside from this, a preliminary examination of the parameter sensitivity is also included.

Based on the study of noisy low resolution spectra generated from our simulation as discussed in $\S$\ref{subsec:SpeDeSN}, we found that adding artificial noise into the QSO spectra increases the amplitude of the flux power spectrum, especially for k $\ge$ 0.01 . This is because random noise and small structures in the flux spectra are indistinguishable. Therefore, we conclude that data in k-space beyond k = 0.01 is untrustworthy for our purposes.

In $\S$\ref{subsec:SenCos}, we provide three sets of simulations with varying $\sigma_{8}$, photoionization parameter $J$, $\Omega_{B}$ to  study parameter sensitivities. We have found that the uncertainty of the photoionization parameter $J$ has a significant influence on the cross-correlation function as well as $\Omega_\Lambda$. Therefore, to constrain the range of $J$ becomes an important goal. In this paper, instead of doing a joint estimation for $J$ and 
$\Omega_\Lambda$, we simply assumed $J$ was known. Estimating $J$ is equivalent to estimating the mean absorption in the Ly$\alpha$ forest which
is sensitive to how one fits the continuum level. We are currently extending our analysis to include uncertainties due to continuum fitting,
and will report on that work in our next paper.

In order to handle the cosmic variance problem, we introduced the MLE technique in the analysis procedures. We then conclude that based on 30 paired QSO spectra with an angular separation of 120", we have 94$\%$ confidence to determine $\Omega_\Lambda$ within 20 $\%$ error ($\epsilon_{20 \%}$), 66$\%$ confidence to determine $\Omega_\Lambda$ within 10 $\%$ error ($\epsilon_{10 \%}$), and 36 $\%$ confidence to determine $\Omega_\Lambda$ within 5 $\%$ error ($\epsilon_{5 \%}$),. Here we have taken finite spectral resolution and noise into account. Our results agree with the results of McDonald (2003) which claims with $N_{pair}$ = 13 $(\theta/1')^{2}$ = 52 can measure $\Omega_\Lambda$ to $\pm$ 0.03 or $\pm$0.04.

\begin{figure}
\plotone{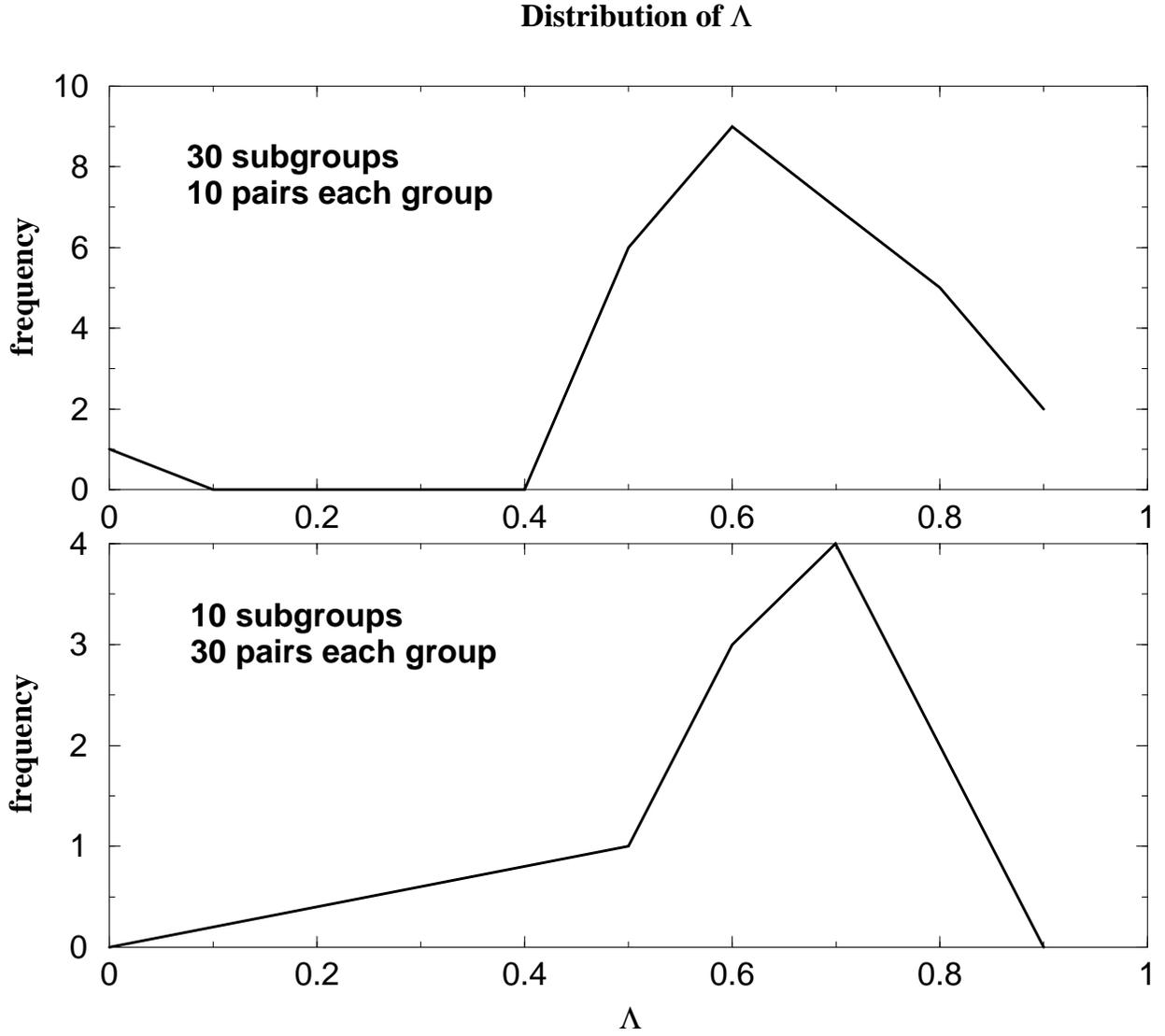}
\caption{The distributions of $\Omega_\Lambda$ based on 10 paired QSO spectra (30 subgroups) and 30 paired QSO spectra (10 subgroups). We have assumed $\Omega_\Lambda$ in the Universe is 0.7. }
\label{fig:Lambda_distri.fig}
\end{figure}

The process of gathering paired QSO spectra is time-consuming. Even before enough QSO pairs have been gathered to constrain the value of $\Omega_\Lambda$, we can achieve a useful result by ruling out unlikely models. Under the assumption that we live in a universe with a high value for $\Omega_\Lambda$, we can rule out the SCDM model at 90 $\%$ confidence by utilizing only 10 pairs of QSO spectra if we assume the the other cosmological parameters are known (Figure (\ref{fig:Lambda_distri.fig})). The tight constraints placed on the other cosmological parameters by WMAP in combination with galaxy clustering data 
structure encourage us that this will be feasible in the near future. The largest remaining uncertainty is the mean level of absorption in the Ly$\alpha$
forest. We are pursing this issue presently and will report our findings in the near future. 

\acknowledgments
We highly appreciate Brian O'Shea, Tridivesh Jena and Pascal Paschos for help editing and organizing the article. We also grateful to the useful discussions with Tridevish Jena and Professor Tytler's group including Nao Suzuki, David Kirkman and John O'Meara. Patrick McDonald deserves special thanks for his ideas during several private communications. This work was supported by NSF grant AST-9803137 under the auspices of the Grand Challenge Cosmology Consortium.


\clearpage

\clearpage
\begin{table*}
\begin{center}
\begin{tabular}{cccccc}
\vspace{0.1cm}
  simulations & $\sigma_{8}$  & $J$ & $\Omega_{B}$ &    h     &$\Omega_\Lambda$  \\
\hline\hline
     A1       &     0.73      &   1.0     &     0.04     &  0.69    &  0.0 (SCDM)      \\
     A2       &     0.73      &   1.0     &     0.04     &  0.69    &  0.5             \\
     A3       &     0.73      &   1.0     &     0.04     &  0.69    &  0.6             \\
A4, B0, C0, D0&     0.73      &   1.0     &     0.04     &  0.69    &  0.7             \\
     A5       &     0.73      &   1.0     &     0.04     &  0.69    &  0.8             \\
     A6       &     0.73      &   1.0     &     0.04     &  0.69    &  0.9             \\
     B1       &     0.657     &   1.0     &     0.04     &  0.69    &  0.7             \\
     B2       &     0.803     &   1.0     &     0.04     &  0.69    &  0.7             \\
     C1       &     0.73      &   0.25    &     0.04     &  0.69    &  0.7             \\
     C2       &     0.73      &   3.0     &     0.04     &  0.69    &  0.7             \\
     D1       &     0.73      &   1.0     &     0.03     &  0.7967  &  0.7             \\
     D2       &     0.73      &   1.0     &     0.05     &  0.617   &  0.7             \\
\hline\hline

\end{tabular}
\vspace{1cm}
\caption{A list of cosmological parameters of our simulations. The varied parameter in Set A is $\Omega_\Lambda$, in Set B is $\sigma_{8}$, in Set C in $J$ and in Set D is $\Omega_{B}$ (equivalent to h).}
\label{table:CosPara.tbl}

\end{center}
\end{table*}
\begin{table*}
\begin{center}
\begin{tabular}{cccccc}
\vspace{0.1cm}
 $\frac{pairs}{group}$ & gp $\#$ & $\Omega_\Lambda$ & P($\epsilon_{20\%}$) & P($\epsilon_{10\%}$) & P($\epsilon_{5\%}$) \\
\hline\hline
 10 & 30 &  0.6 $\pm$ 0.15 & 56 $\%$ & 30 $\%$ & 15 $\%$ \\
 15 & 20 &  0.6 $\pm$ 0.1  & 64 $\%$ & 34 $\%$ & 17 $\%$ \\
 20 & 15 &  0.64$\pm$ 0.1  & 77 $\%$ & 44 $\%$ & 23 $\%$ \\
 25 & 12 &  0.7 $\pm$ 0.1  & 80 $\%$ & 48 $\%$ & 25 $\%$ \\
 30 & 10 &  0.7 $\pm$ 0.07 & 94 $\%$ & 66 $\%$ & 36 $\%$ \\
\hline\hline

\end{tabular}
\vspace{1cm}
\caption{Probability of getting $\Lambda$ within 20$\%$, 10$\%$ and 5$\%$ error. Each row in the table indicates one statistical study. The first column is the number of QSO pairs in one subgroup while the second column is the number of subgroups in each study. The third column denotes the result of $\Omega_\Lambda$. Finally, the fourth, fifth, and the sixth columns are the probability of getting $\Omega_\Lambda$ within 20$\%$, 10$\%$ and 5$\%$ error.}
\label{table:Lambda_Prob.tbl}

\end{center}
\end{table*}
\clearpage
\appendix
\chapter{A}
\section{Maximum-Likelihood Estimation}
\label{sec:MLE}
  For a random sample $x_{1}$, $x_{2}$, ......$x_{n}$ from a population with a probability density function (PDF) depending on a parameter $\theta$, the function t = t($x_{1}$, $x_{2}$, ...... $x_{n}$) is called a "statistic" if it does not depend on any other unknown parameters. Then the term "estimator" denotes a function, method or prescription used to find a value of an unknown parameter. In general, t is an estimator of the unknown $\theta$. For a continuous or discrete population which has probability density function f(x $\mid$ $\theta$), the likelihood of n observations $x_{1}$, $x_{2}$, ...... $x_{n}$ for a specific $\theta$ is given by

\begin{equation}
  L(x_{1},x_{2},...,x_{n} \mid \theta) = \Pi_{i=1}^{n} f(x_{i} \mid \theta)
\label{eqn:Likelihood_def.eqn}
\end{equation}

We can think of $L(x_{1},x_{2},...,x_{n} \mid \theta)$ as a function of $\theta$ and call it the "likelihood function", denoted by $L$ or $L(\underline{x} \mid \theta)$. To be more general, if we consider n independent experiments with the same physical parameter $\theta$ from these n sets, $\underline{x}_{1}$, $\underline{x}_{2}$, \ldots, $\underline{x}_{n}$, of observations, the corresponding likelihood functions of each set are L($\underline{x}_{1}$ $\mid$ $\theta$), L($\underline{x}_{2}$ $\mid$ $\theta$),\ldots, and L($\underline{x}_{n}$ $\mid$ $\theta$), respectively. Then the combined-likelihood of all observations is :

\begin{equation}
  L(\underline{{x}_{1}}, \underline{{x}_{2}},\ldots ,\underline{{x}_{n}} \mid \theta) \equiv  \Pi_{i1} f_{1}(x_{i1} \mid \theta) \Pi_{i2} f_{2}(x_{i2} \mid \theta) ..........  \Pi_{in} f_{n}(x_{in} \mid \theta)
\label{eqn:L_joint.eqn}
\end{equation}

which is equivalent to

\begin{equation}
  log( L(\underline{{x}_{1}}, \underline{{x}_{2}}, \ldots , \underline{{x}_{n}} \mid \theta) \equiv log(f_{1}(x_{i1} \mid \theta)) + \sum_{i2} log(f_{2}(x_{i2} \mid \theta)) +  ..........+  \sum_{in} log(f_{n}(x_{in} \mid \theta))
\label{eqn:Log_L_joint.eqn}
\end{equation}

  Maximizing the combined likelihood in Equation (\ref{eqn:L_joint.eqn}) or its equivalent expression in Equation (\ref{eqn:Log_L_joint.eqn}) gives us an estimation of $\theta$ denoted by $\hat{\theta}$ (Frodesen et al. 1978). In this work, the unknown parameter $\theta$ is chosen to be $\Omega_\Lambda$. Each $x_i$ is a data point in the corresponding cross-correlation function. The probability density function $f_i(x_i \mid \theta)$ is the PDF of the cross-correlation at a velocity separation for a given $\Omega_\Lambda$ while other parameters are fixed (Details are described in $\S$\ref{subsec:CreatPDF} and $\S$\ref{subsec:Likelihood_Cal}).

\vspace{1.5in}

\chapter{B}
\section{Error Propagation}
\label{subsec:ErrPro}
 Let $E[f(x_0)]$ denote the error of any given function f(x) at $x_0$. As mentioned in $\S$\ref{subsec:CrossCorre}, the mean flux $\bar{f}$ of a paired QSO spectra:
\begin{equation}
\bar{f} = \frac{ \sum_{i=1}^{n} (f_{1}(z_i) + f_2(z_i))}{2n}
\end{equation}
where n is the number of points in a spectrum, $f_{1}(z_i)$ and  $f_{2}(z_i)$, i=1, 2, $\ldots$ n are the data points from the first and second flux spectra of a QSO pair. Then based on the error propagation formulas (Harrison, 2001) we have
\begin{equation}
E[\bar{f}] = \frac{ \sqrt{ \sum_{i=1}^{n} (E[f_{1}(z_i)]^2 + E[f_2(z_i)]^2)}}{2n}
\end{equation}
where $E[f_{1}(z_i)]$ is the error of the point $f_{1}(z_i)$ and $E[f_{2}(z_i)]$ is the error of the point $f_{2}(z_i)$. For the overflux $\delta$ f = $\frac{f-\bar{f}}{\bar{f}}$, we have E[$f -\bar{f}$] = $\sqrt{E[f]^2 + E[\bar{f}]^2}$. So
\begin{eqnarray}
 E(\delta f) = E[ \frac{f-\bar{f}}{\bar{f}} ] = \delta f \sqrt{\frac{E[f -\bar{f}]^2}{ (f-\bar{f})^2 } + (\frac{E[\bar{f}]}{\bar{f}})^2 }  = \delta f \sqrt{\frac{E[f]^2 + E[\bar f]^2}{ (f-\bar{f})^2 } + (\frac{E[\bar{f}]}{\bar{f}})^2 }
\end{eqnarray}

From $\S$\ref{subsec:CrossCorre}, $\xi(\Delta v_{\parallel}) = <\delta_{f1}(z_{i}) \delta_{f2}(z_{j})>$, where the average is over all possible (i,j) permutations. We consquently get

\begin{equation}
E[\delta_{f1} (z_i) \delta_{f2} (z_j)] = \delta_{f1}(z_i)\delta_{f1}(z_i) \sqrt{(\frac{E[\delta_{f1}(z_i)]}{\delta_{f1}(z_i)})^2 +(\frac{E[\delta_{f2}(z_j)]}{\delta_{f2}(z_j)})^2}
\end{equation}

Therefore,
\begin{equation}
E[\xi(\Delta v_{\parallel})] = \frac{1}{N} \sqrt{\sum_{i,j} (\frac{E[\delta_{f1} (z_i) \delta_{f2} (z_j)]}{\delta_{f1} (z_i) \delta_{f2} (z_j)})^2}
\end{equation}

\clearpage



\end{document}